\documentclass[a4paper,11pt]{article}
\pdfoutput=1 % if your are submitting a pdflatex (i.e. if you have
\usepackage{jcappub} % for details on the use of the package, please
\usepackage[T1]{fontenc} % if needed

\usepackage{subfigure}
\usepackage{enumitem}
\usepackage{pifont}
\setlist[itemize]{itemsep=0.5ex,parsep=0pt,label=\checkmark}

\usepackage{color, colortbl}
\usepackage[dvipsnames]{xcolor}
\definecolor{Gray}{gray}{0.9}
\definecolor{Gray2}{gray}{0.95}
\definecolor{LightCyan}{rgb}{0.88,1,1}

\usepackage{subfigure}

\usepackage{cancel}

\usepackage[normalem]{ulem}

\title{\boldmath Correlations for an anisotropic polarized stochastic gravitational wave background in pulsar timing arrays}

\author[a,b,1]{Reginald Christian Bernardo\note{Corresponding author}}
\author[c]{, Guo-Chin Liu}
\author[b,d]{, Kin-Wang Ng}
\affiliation[a]{Asia Pacific Center for Theoretical Physics, \\ Pohang 37673, Republic of Korea}
\affiliation[b]{Institute of Physics, Academia Sinica, \\ Taipei 11529, Taiwan}
\affiliation[c]{Department of Physics, Tamkang University, Tamsui, \\ New Taipei 25137, Taiwan}
\affiliation[d]{Institute of Physics, Academia Sinica, \\ Taipei 11529, Taiwan}
\emailAdd{reginald.bernardo@apctp.org}
\emailAdd{gcliu@gms.tku.edu.tw}
\emailAdd{nkw@phys.sinica.edu.tw}
\abstract{
The recent compelling observation of the nanohertz stochastic gravitational wave background has brought to light a new galactic arena to test gravity. In this paper, we derive a formula for the most general expression of the stochastic gravitational wave background correlation that could be tested with pulsar timing and future square kilometer arrays. Our expressions extends the harmonic space analysis, also often referred to as the power spectrum approach, to predict the correlation signatures of an anisotropic polarized stochastic gravitational wave background with subluminal tensor, vector, and scalar gravitational degrees of freedom. We present the first few nontrivial anisotropy and polarization signatures in the correlation and discuss their dependence on the gravitational wave speed and pulsar distances. Our results set up tests that could potentially be used to rigorously examine the isotropy of the stochastic gravitational wave background and strengthen the existing constraints on possible non-Einsteinian polarizations in the nanohertz gravitational wave regime.
}

\begin{document}
\maketitle
\flushbottom

\section{Introduction}
\label{sec:introduction}

The recent compelling evidence of pulsar timing array (PTA) groups of the Hellings \& Downs (HD) \cite{Hellings:1983fr, Jenet:2014bea} curve is an astronomical milestone \cite{NANOGrav:2023gor, Reardon:2023gzh, EPTA:2023sfo, Xu:2023wog}. First off, the observation highlights gravitational waves in a stochastic superposition, producing the renowned quadrupolar correlation pattern and complementing the single-source-picture provided by ground-based gravitational wave (GW) detectors \cite{LIGOScientific:2021djp, Romano:2023zhb}. Sustaining decades of effort of precisely timing millisecond pulsars over the Galaxy is no small feat  \cite{Detweiler:1979wn, Romano:2016dpx, Burke-Spolaor:2018bvk, NANOGrav:2020spf}. This has opened up a new window to challenge standard models of early universe cosmology \cite{Chen:2019xse, NANOGrav:2021flc, Xue:2021gyq, Moore:2021ibq, NANOGrav:2023hvm, EPTA:2023xxk, Vagnozzi:2023lwo, Franciolini:2023wjm, Bai:2023cqj, Megias:2023kiy, Jiang:2023gfe, Zhang:2023nrs, Figueroa:2023zhu, Bian:2023dnv, Niu:2023bsr, Depta:2023qst, Abe:2023yrw, Servant:2023mwt, Ellis:2023oxs, Bhaumik:2023wmw, Ahmed:2023pjl, Antusch:2023zjk, Aghaie:2023lan, Konoplya:2023fmh, Basilakos:2023xof, Basilakos:2023jvp, Ben-Dayan:2023lwd, Ahmadvand:2023lpp, Choudhury:2023hfm} and a debate whether the main source of the observed stochastic gravitational wave background (SGWB) is astrophysical supermassive black hole binaries \cite{Rajagopal:1994zj, Phinney:2001di, Sesana:2008mz, Ravi:2014nua, Shannon:2015ect, Mingarelli:2017fbe, Liu:2021ytq, NANOGrav:2021ini, NANOGrav:2023pdq, EPTA:2023gyr, Ellis:2023dgf, Cannizzaro:2023mgc}. In the same vein, a promising PTA frontier for gravitational physics in the nanohertz GW regime remains to be further advocated, despite a couple of papers \cite{Chamberlin:2011ev, Gong:2018vbo, Qin:2020hfy, Liang:2021bct, Bernardo:2022rif, Wu:2023pbt, Liang:2023ary, Bernardo:2023pwt} hinting at many interesting phenomena to be explored in this regime of Galactic size GWs.

From a gravitational physics standpoint, in the isotropic case, which is the leading signal expected from the SGWB, it has been very well established that the HD curve resulting from the Earth term of the cross-correlation of pulsar-pair timing residuals is an excellent approximation at large scales. Significant departures to this at small scales could however be traced to the often-dropped pulsar term modulations \cite{Mingarelli:2014xfa, Ng:2021waj}. Furthermore, pulling away from the mainstream picture of gravity that is general relativity (GR), it turns out that an exciting variety of correlation patterns could be realized and potentially tested with data \cite{Bernardo:2023mxc, Bernardo:2023zna, Chen:2023uiz, Bi:2023ewq}. Theories that admit a nontrivial dispersion of GWs in the nanohertz regime are particularly interesting, especially if it could somehow also manage to satisfy the strong multimessenger bound on the GW speed in the sub-kilohertz band \cite{LIGOScientific:2017zic, LIGOScientific:2021sio}.

A most important progress in this direction is the realization of the cosmic variance for GR \cite{Allen:2022dzg, Allen:2022ksj, Caliskan:2023cqm} and alternative theories of gravity \cite{Bernardo:2022xzl, Bernardo:2023bqx}. Analogous to the way this is played in CMB science and cosmology \cite{Srednicki:1993ix, Roebber:2016jzl, Gair:2014rwa}, the cosmic variance arises from the fact that measurements can only be done in one universe and so is invaluable to consider when comparing the predictions of gravity theories. Based on this, a quadrupolar SGWB can be singled out to come from tensorial gravitational degrees of freedom (d.o.f.s), which could be luminal or subluminal for the meantime for nanohertz GWs. Inevitably, the question of `how quadrupolar the SGWB is?' comes up---which is synonymous to testing nanohertz gravity. The correlation signatures from vector and scalar GWs have similarly been understood together with their cosmic variances \cite{Bernardo:2022xzl, Bernardo:2023bqx}.

All of the above important results hinge on one property that the SGWB is isotropic. However, this could be tested more rigorously by looking for departures from this isotropic assumption, such as when the SGWB has anisotropic and polarized components. This was discussed in a handful of trailblazing papers \cite{Mingarelli:2013dsa, Taylor:2013esa, Gair:2014rwa, Kato:2015bye, Chu:2021krj, Sato-Polito:2021efu, Liu:2022skj, NANOGrav:2023tcn} that pave the road for possible breakthrough discovery of anisotropy and polarization in the SGWB. Most recent results touch even further the realms of utilizing anisotropy and polarization in the SGWB to detect non-Einsteinian GW polarizations \cite{Tasinato:2023zcg, AnilKumar:2023yfw, AnilKumar:2023kvt}.

In Ref. \cite{Liu:2022skj}, it has been shown that the correlation for an anisotropic polarized SGWB with luminal tensor modes can be teased out using special functions traceable to the rotational properties of the fundamental field. This laid out the groundwork for a fast and efficient algorithm for calculating the correlations of an anisotropic polarized SGWB. On the other hand, in Ref. \cite{AnilKumar:2023yfw}, an impressive compactified expression for the correlation for luminal GWs including non-Einsteinian polarizations was given in terms of bipolar spherical harmonics (see Section \ref{sec:changeofbasis} for more details). This work shares the theme of Ref. \cite{AnilKumar:2023yfw}\footnote{Ref. \cite{AnilKumar:2023yfw} appeared in the arXiv when we were wrapping up this paper.} for going beyond standard luminal tensor modes and builds on the language of Ref. \cite{Liu:2022skj}, thus keeping pulsar terms and utilizing an established numerical routine, to setup a formalism for computing the anisotropic and polarized components of the SGWB for {\it subluminal} tensor, vector, and scalar gravitational d.o.f.s {\it and} pulsars at finite distances. Section \ref{sec:summary} summarizes our main results into readily codable formula for computing the various correlation components of the SGWB.

The rest of this paper proceeds as follows. In Section \ref{sec:pulsartimingobservables}, we setup the formalism for looking at the pulsar timing main observable, the timing residue, and in Section \ref{sec:correlationforanisotropies}, we calculate its two-point correlation function, obtaining what we perceive as generalized expressions for the inter-pulsar correlation with anisotropy and polarization. Putting this to good use, in Section \ref{sec:discussion}, we present and discuss the first nontrivial correlation signatures of anisotropy and polarization for tensor, vector, and scalar GWs. In Section \ref{sec:changeofbasis}, we briefly touch on differences between our approach and that by Ref. \cite{AnilKumar:2023yfw}. We wrap up the paper in Section \ref{sec:conclusions}.

In Appendix \ref{sec:orfs_review}, we briefly review useful analytical expressions for the correlation of an isotropic SGWB. In Appendix \ref{sec:3Y3j}, we provide useful relations and identities for the spin-weighted spherical harmonics and the Wigner-3j symbol, and in Appendices \ref{sec:gwpolarizations}-\ref{sec:Jlmcalculations} we express the basis tensors for Einstein and non-Einsteinian GW polarizations and give the full detailed derivation of the generalized projection factors.

\section{Summary of main results}
\label{sec:summary}

We summarize the formulae derived in the bulk of this paper (Sections \ref{sec:pulsartimingobservables}-\ref{sec:correlationforanisotropies} and Appendix \ref{sec:Jlmcalculations}) for the correlations of an anisotropic polarized SGWB. These are implemented in the public code \href{https://github.com/reggiebernardo/PTAfast}{PTAfast} \cite{2022ascl.soft11001B} with ready-to-use modules for generating the components of a generally anisotropic and polarized SGWB.

For tensor GWs with frequency $f$, the SGWB-induced spatial correlations between two pulsars $a$ and $b$ with distances $D_a$ and $D_b$ are given by
\begin{equation}
\label{eq:gammaIV_tensor_summary}
\begin{split}
    \gamma_{lm}^{I,V}\left( f D_a, fD_b, \zeta \right) = \sum_{l_1 l_2}
    & (-1)^m \left( \dfrac{2 l_1 + 1}{4\pi} \right) \left[ 1 \pm (-1)^{l + l_1 + l_2} \right] C_{l_1 l_2}^{\rm \bf T}(fD_a, fD_b) Y_{l_2 m}\left( \zeta, 0 \right) \\
    & \times \sqrt{(2l+1)(2l_2 + 1)}
    \left( \begin{array}{ccc}
    l & l_1 & l_2 \\
    0 & -2 & 2
    \end{array} \right)
    \left( \begin{array}{ccc}
    l & l_1 & l_2 \\
    m & 0 & -m
    \end{array} \right)
\end{split}
\end{equation}
and
\begin{equation}
\label{eq:gammaQU_tensor_summary}
\begin{split}
    \gamma_{lm}^{Q \pm i U}\left( f D_a, fD_b, \zeta \right) = \sum_{l_1 l_2}
    & (-1)^m \left( \dfrac{2l_1+1}{4\pi} \right) C_{l_1 l_2}^{\rm \bf T}(fD_a, fD_b) Y_{l_2 m}\left( \zeta, 0 \right) \\
    & \times \sqrt{ (2l+1)(2l_2 + 1) }
    \left( \begin{array}{ccc}
    l & l_1 & l_2 \\
    \mp 4 & \pm 2 & \pm 2
    \end{array} \right)
    \left( \begin{array}{ccc}
    l & l_1 & l_2 \\
    m & 0 & -m
    \end{array} \right) \,,
\end{split}
\end{equation}
where the superscripts $I, V, Q, U$ stand for Stokes parameters, e.g., an isotropic unpolarized SGWB would have $I \neq 0$ and $V = Q = U = 0$, or in terms of the correlations $\gamma_{00}^I(\zeta) \neq 0$ and $\gamma_{00}^V(\zeta) = \gamma_{00}^{Q \pm i U}(\zeta) = 0$. The $\left(\begin{array}{ccc} a & b & c \\ d & e & f \end{array}\right)$'s are Wigner-3j symbols (Appendix \ref{sec:3Y3j}). The $C_{l_1 l_2}^{\rm \bf T}(fD_a, fD_b)$'s are a generalization of the correlation power spectrum multipoles or harmonic space coefficients \cite{Gair:2014rwa, Qin:2018yhy, Qin:2020hfy, Ng:2021waj, Liu:2022skj, Bernardo:2022rif} and defined as
\begin{equation}
\label{eq:Cl_T_summary}
    C_{l_1 l_2}^{\rm \bf T}(fD_a, fD_b) = \dfrac{J^{\rm \bf T}_{l_1}\left( fD_a \right) J_{l_2}^{{\rm \bf T}*}\left( f D_b \right)}{\sqrt{\pi}}
\end{equation}
where the $J_l^{\rm \bf T}(x)$'s are given by
\begin{equation}
\label{eq:Jl_T_summary}
    J^{\rm \bf T}_l\left(fD\right) = \sqrt{2} \pi i^l \sqrt{\dfrac{(l + 2)!}{(l - 2)!}} \int_0^{2\pi fDv} \dfrac{dx}{v} \ e^{ix/v} \dfrac{j_l(x)}{x^2} \,,
\end{equation}
with the $j_l(x)$'s being spherical Bessel functions of the first kind. The isotropic component ($l = m = 0$) gives the HD curve for luminal GWs, $v = 1$, and infinite pulsar distances, $D_{a,b} \rightarrow \infty$. The quantity $J_l(x)$'s can be regarded as a generalization to the projection factors \cite{Qin:2018yhy, Qin:2020hfy} that are characteristic of the different GW polarizations (Appendix \ref{sec:gwpolarizations}).

For vector GWs, the correlations turn out to be
\begin{equation}
\label{eq:gammaIV_vector_summary}
\begin{split}
    \gamma_{lm}^{I,V}\left( f D_a, fD_b, \zeta \right) = \sum_{l_1 l_2}
    & (-1)^{1 + m} \left( \dfrac{2l_1 + 1}{4\pi} \right) \left[ 1 \pm (-1)^{l + l_1 + l_2} \right] C_{l_1 l_2}^{\rm \bf V}(fD_a, fD_b) Y_{l_2 m}\left( \zeta, 0 \right) \\
    & \sqrt{ (2l+1)(2l_2 + 1) }
    \left( \begin{array}{ccc}
    l & l_1 & l_2 \\
    0 & -1 & 1
    \end{array} \right)
    \left( \begin{array}{ccc}
    l & l_1 & l_2 \\
    m & 0 & -m
    \end{array} \right)
\end{split}
\end{equation}
and
\begin{equation}
\label{eq:gammaQU_vector_summary}
\begin{split}
    \gamma_{lm}^{Q \pm i U}\left( f D_a, fD_b, \zeta \right) = \sum_{l_1 l_2}
    & (-1)^m \left( \dfrac{2l_1 + 1}{4\pi} \right) C_{l_1 l_2}^{\rm \bf V}(fD_a, fD_b) Y_{l_2 m}\left( \zeta, 0 \right)  \\
    & \sqrt{ (2l+1)(2l_2 + 1) }
    \left( \begin{array}{ccc}
    l & l_1 & l_2 \\
    \mp 2 & \pm 1 & \pm 1
    \end{array} \right)
    \left( \begin{array}{ccc}
    l & l_1 & l_2 \\
    m & 0 & -m
    \end{array} \right) \,,
\end{split}
\end{equation}
where the $C_{l_1 l_2}^{\rm \bf V}(fD_a, fD_b)$'s are the vector GW correlation power spectrum multipoles defined by
\begin{equation}
\label{eq:Cl_V_summary}
    C_{l_1 l_2}^{\rm \bf V}(fD_a, fD_b) = \dfrac{J^{\rm \bf V}_{l_1}\left( fD_a \right) J_{l_2}^{{\rm \bf V}*}\left( f D_b \right)}{\sqrt{\pi}}
\end{equation}
and
\begin{equation}
\label{eq:Jl_V_summary}
    J^{\rm \bf V}_l\left(fD\right) = 2 \sqrt{2} \pi i^l \sqrt{l(l+1)} \int_0^{2\pi fDv} \dfrac{dx}{v} \ e^{ix/v} \dfrac{d}{dx} \left( \dfrac{j_l(x)}{x} \right) \,.
\end{equation}
The notable difference between the tensor and vector cases appear in the Wigner-3j symbols, indicative of the spin of the underlying gravitational field.

For scalar GWs, the correlations are given by
\begin{equation}
\label{eq:gamma_scalar_summary}
\begin{split}
    \gamma_{lm} \left( fD_a, fD_b, \zeta \right) = \sum_{l_1 l_2}
    & (-1)^m \left( \dfrac{2l_1 + 1}{4\pi} \right) C_{l_1 l_2}^{\rm \bf S}(fD_a, fD_b) Y_{l_2 m}\left( \zeta, 0 \right) \\
    & \sqrt{ (2l+1)(2l_2 + 1) }
    \left( \begin{array}{ccc}
    l & l_1 & l_2 \\
    0 & 0 & 0
    \end{array} \right)
    \left( \begin{array}{ccc}
    l & l_1 & l_2 \\
    m & 0 & -m
    \end{array} \right) \,,
\end{split}
\end{equation}
where $C_{l_1 l_2}^{\rm \bf S}(fD_a, fD_b)$'s are the scalar GW correlation power spectrum multipoles given by
\begin{equation}
    C_{l_1 l_2}^{\rm \bf S}(fD_a, fD_b) = 32 \pi^2 J^{\rm \bf S}_{l_1}\left( fD_a \right) J_{l_2}^{{\rm \bf S} *}\left( f D_b \right) \,
\end{equation}
and
\begin{equation}
\label{eq:Jls_scalar_summary}
    {J}^{\rm \bf S}_l \left(fD\right) = - \dfrac{i^l}{2} \int_0^{2\pi fDv} \dfrac{dx}{v} e^{ix/v} \left( -\dfrac{\left( j_l''(x) + j_l(x) \right)}{\sqrt{2}} + \dfrac{1-v^2}{\sqrt{2}} j_l''(x)  \right) \,.
\end{equation}
The decomposition in Stokes parameters is not utilized for scalar GWs since scalar modes anchor a purely longitudinal GW polarization. As with previous cases, the Wigner-3j symbol in the correlation gives away the spin-$0$ nature of the scalar field. It is useful to note that the $J_l^{\rm \bf S}(x)$ integral for scalar GWs reduces to a total boundary that completely cancels out the contribution for $l \geq 2$ at infinite pulsar distances \cite{Qin:2020hfy, Bernardo:2022vlj, Bernardo:2023pwt}. The two terms in the integrand of \eqref{eq:Jls_scalar_summary} were separated only to tease out the transverse and longitudinal components; the first term, $\propto j''(x) + j(x)$, is the transverse (`breathing') mode contribution and the second term, $\propto (1-v^2)$, is the scalar longitudinal mode contribution. In general, this implies that correlations due to scalar GWs are physically distinguished by pronounced monopolar and dipolar components.

\section{Pulsar timing observables and correlation}
\label{sec:pulsartimingobservables}

In this section, we review the timing residue in the context of a passing gravitational wave and the pulsar timing array correlation.

\subsection{Nanohertz GWs and the Timing Residue}
\label{subsec:nanohertzgwsandthetimingresidue}

Consider a plane gravitational wave with a frequency $f$, wave vector $\hat{k}$, and polarization $A$. In this case, `polarization' means the various independent ways a gravitational wave displaces masses on its path: scalar (breathing, longitudinal), vector (x, y), tensor $(+, \times)$ \cite{Chamberlin:2011ev}. We can represent a general gravitational wave as a superposition of these plane waves,
\begin{equation}
\label{eq:gw_general}
    h_{ij}\left(\eta, \vec{x}\right) = \sum_A \int_{-\infty}^\infty df \int_{S^2} d\hat{k} \ h_A\left(f, \hat{k}\right) \varepsilon_{ij}^A e^{-2\pi i f \left( \eta - v \hat{k} \cdot \vec{x} \right)} \,,
\end{equation}
where $\varepsilon_{ij}^A$ are basis polarization tensors (Appendix \ref{sec:gwpolarizations}) and $v = d\omega/dk$ is the wave's group velocity. In this section, we establish the connection between a gravitational wave $h_{ij}$ and pulsar timing.

The relevant observable is the pulsar timing residue, $r(t)$, which is the remainder of a pulsar's time-of-arrival radio signals after subtracting away effects due to known pulsar astrophysics and systematics. Thus, in principle, a nonvanishing pulsar timing residue can be attributed to external forces. In this work, we assume that gravity is mainly responsible for nontrivial pulsar timing residuals and derive the inter-pulsar correlation induced by a stochastic gravitational wave background.

We begin by writing the pulsar timing residual as \cite{Mingarelli:2013dsa, Qin:2018yhy, Ng:2021waj}
\begin{equation}
\label{eq:timing_residual}
    r\left(t\right) = \int_0^t dt' \ z\left(t'\right) \,,
\end{equation}
where $z(t)$ is a redshift space fluctuation induced by a passing gravitational wave, $h_{ij}\left(\eta, \vec{x}\right)$. If a photon were emitted at time $\eta_e$ and received (by a detector) at a later time $\eta_r > \eta_e$, then the redshift space fluctuation is given by
\begin{equation}
    z\left(t, \hat{e}\right) = - \dfrac{1}{2} \int_{t + \eta_e}^{t + \eta_r} d\eta \ d^{ij} \partial_\eta h_{ij} \left( \eta, \vec{x} \right) \,,
\end{equation}
where $d^{ij}$ is the detector tensor, $d^{ij} = e^i e^j$, and $\hat{e}$ is a unit vector pointing from Earth toward a pulsar. By substituting the GW \eqref{eq:gw_general} into the expression of the pulsar timing residue, we obtain 
\begin{equation}
\begin{split}
    r\left(t, \hat{e}\right) = & \int_0^t dt' \left( - \dfrac{1}{2} \right) \int_{t' + \eta_e}^{t' + \eta_r} d \eta \ d^{ij} \sum_{A} \int_{-\infty}^\infty df \int_{S^2} d\hat{k} \ h_A \left(f, \hat{k}\right) \varepsilon_{ij}^A\left(\hat{k}\right) \\
    & \ \ \times \left(-2\pi i f\right) e^{-2\pi i f \eta} \ 4\pi \sum_{lm} i^l j_l\left(2\pi fv \left( t' + \eta_r - \eta \right) \right) Y_{lm}^*\left(\hat{k}\right) Y_{lm}\left(\hat{e}\right) \,,
\end{split}
\end{equation}
where $\vec{x} = |\vec{x}| \hat{e} = \left( t' + \eta_r - \eta \right) \hat{e}$ is the position vector to the pulsar at time $t'$ and $j_l(x)$ is the spherical Bessel function of the first kind. Integrating over the time variables $t'$ and $\eta$, we obtain our final simplified expression of the pulsar timing residual,
\begin{equation}
\label{eq:timingresidueharmonics_gw}
\begin{split}
    r\left(t,\hat{e}\right) = \ & 2 \pi \sum_A \int_{-\infty}^\infty df \int_{S^2} d\hat{k} \ \left( 1 - e^{-2\pi i f t} \right) \left( \dfrac{e^{-2\pi i f \eta_r}}{2\pi f} \right) \\
    & \ \ \ \ \times h_A\left(f, \hat{k}\right) \int_0^{2\pi f Dv} \dfrac{dx}{v} e^{ix/v} \left[ d^{ij} \varepsilon_{ij}^A \right] \sum_{lm} i^l j_l\left(x\right) Y^*_{lm}\left(\hat{k}\right) Y_{lm}\left(\hat{e}\right) \,.
\end{split}
\end{equation}
We use this later to calculate the two-point function of the GW correlation.

\subsection{The Timing Residual Rower Spectrum}
\label{subsec:gwcorrelationpowerspectrum}

In this section, we derive the two-point function of the timing residuals of a pair of pulsars. This important quantity contains the crucial information that singles out GW phenomena.

We proceed by expanding the timing residue in spherical harmonics, a.k.a. harmonic analysis,
\begin{equation}
\label{eq:timingresidueharmonics_general}
    r\left(t, \hat{e}\right) = \sum_{lm} a_{lm} Y_{lm} \left( \hat{e} \right) \,.
\end{equation}
The coefficients $a_{lm}$'s for a passing GW can be read by comparing \eqref{eq:timingresidueharmonics_gw} and \eqref{eq:timingresidueharmonics_general}. In the presence of an ensemble of GWs, i.e., a stochastic gravitational wave background, the two-point function of the timing residuals can be shown to be
\begin{equation}
    \langle r\left(t_a, \hat{e}_a\right) r\left(t_b, \hat{e}_b\right) \rangle = \sum_{l_1, m_1} \sum_{l_2, m_2} \langle a_{l_1 m_1} a^*_{l_2 m_2} \rangle Y_{l_1 m_1}\left( \hat{e}_a \right) Y^*_{l_2 m_2}\left( \hat{e}_b \right)
\end{equation}
where
\begin{equation}
\label{eq:two_point_lm}
\begin{split}
    \langle a_{l_1 m_1} a^*_{l_2 m_2} \rangle = & \int_{-\infty}^\infty \dfrac{df}{\left(2\pi f\right)^2} \left( 1 - e^{-2\pi i f t_a} \right) \left( 1 - e^{2\pi i f t_b} \right) \\
    & \ \ \times \sum_{A_1, A_2} \int_{S^2} d\hat{k} \ P_{A_1 A_2}\left( f, \hat{k} \right) J^{A_1}_{l_1 m_1} \left( f D_a, \hat{k} \right) J^{A_2 *}_{l_2 m_2} \left( f D_b, \hat{k} \right)
\end{split}
\end{equation}
and
\begin{equation}
\label{eq:Jlm_def}
    J_{lm}^A \left( fD, \hat{k} \right) = \int_0^{2\pi f D v} \dfrac{d x}{v} \ e^{i x/v} \sum_{LM} 2 \pi i^L Y^*_{LM} \left( \hat{k} \right) j_L(x) \int_{S^2} d\hat{e} \ d^{ij} \varepsilon_{ij}^A\left(\hat{k}\right) Y_{LM}\left( \hat{e} \right) Y_{lm}^*\left(\hat{e}\right) \,.
\end{equation}
In \eqref{eq:two_point_lm} and \eqref{eq:Jlm_def}, $P_{A_1 A_2} \left(f, \hat{k}\right)$ is the amplitude of the GW two-point function,
\begin{equation}
    \langle h_A\left(f, \hat{k}\right) h_B^* \left( f', \hat{k}'\right) \rangle = \delta \left( f - f' \right) \delta \left( \hat{k} - \hat{k}' \right) P_{AB} \left(f, \hat{k}\right) \,.
\end{equation}
In the isotropic case, $P_{AB} \left(f, \hat{k}\right) = \delta_{AB} {\cal P}(f)$, where ${\cal P}(f)$ is related to the GW density, often expressed as $\Omega_{\rm GW} h^2$. This quantity is of significant interest for distinguishing early universe models against the vanilla astrophysical SMBHBs as the dominant source of the SGWB.

Our main interest on the other hand is the overlap reduction function (ORF), measuring the angular correlation between pulsar pairs. In harmonic space, following the formalism outlined, this can be shown to be
\begin{equation}
\label{eq:orf_general}
\begin{split}
    \gamma_{lm}^A \left( fD_a, fD_b; \hat{e}_a, \hat{e}_b \right) =
    & \sum_{l_1, m_1} \sum_{l_2, m_2} Y_{l_1 m_1}\left( \hat{e}_a \right) Y^*_{l_2 m_2} \left( \hat{e}_b \right) \\
    & \ \ \times \int_{S^2} d\hat{k} \ Y_{lm}\left(\hat{k}\right) J^A_{l_1 m_1} \left( f D_a, \hat{k} \right) J^{A*}_{l_2 m_2} \left( f D_b, \hat{k} \right) \,.
\end{split}
\end{equation}

In \cite{Bernardo:2022rif}, we were only concerned with an isotropic SGWB; in other words, the $l = m = 0$ limit of the above general expression. In this paper, we consider the general case of an anisotropic polarized SGWB. For this purpose, we define for transverse modes right- and left-handed helicity basis tensors ($\varepsilon^{R/L}$) and amplitudes ($h_R, h_L$), and their associated Stokes parameters $I, Q, U$, and $V$ as
\begin{eqnarray}
I &=& \left( \langle h_R h_R^* \rangle + \langle h_L h_L^* \rangle \right)/2 \\
Q + i U &=& \langle h_L h_R^* \rangle \\
Q - i U &=& \langle h_R h_L^* \rangle \\
V &=& \left( \langle h_R h_R^* \rangle - \langle h_L h_L^* \rangle \right)/2 \,.
\end{eqnarray}
In this case, we are able to write down the correlation associated with anisotropy and polarization as \cite{Liu:2022skj}
\begin{equation}
\begin{split}
\label{eq:orf_IV}
    \gamma_{lm}^{I,V}\left( fD_a, fD_b; \hat{e}_a, \hat{e}_b \right) = 
    & \sum_{l_1, m_1} \sum_{l_2, m_2} Y_{l_1 m_1}\left(\hat{e}_a\right) Y^*_{l_2 m_2}\left(\hat{e}_b\right) \\
    & \ \ \times \int_{S^2} d\hat{k} \ Y_{lm}\left(\hat{k}\right) \mathbb{J}^{I,V}_{l_1 m_1 l_2 m_2} \left( fD_a, fD_b, \hat{k} \right)
\end{split}
\end{equation}
and
\begin{equation}
\label{eq:orf_QU}
\begin{split}
    \gamma_{lm}^{Q \pm i U}\left( fD_a, fD_b; \hat{e}_a, \hat{e}_b \right) =
    & \sum_{l_1, m_1} \sum_{l_2, m_2} Y_{l_1 m_1}\left(\hat{e}_a\right) Y^*_{l_2 m_2}\left(\hat{e}_b\right) \\
    & \ \ \times \int_{S^2} d\hat{k} \ Y_{lm}\left(\hat{k}\right) \mathbb{J}^{Q \pm i U}_{l_1 m_1 l_2 m_2} \left( fD_a, fD_b, \hat{k} \right) \,,
\end{split}
\end{equation}
where the functions $\mathbb{J}^{X}_{lm}(d_a, d_b, \hat{k})$'s are given by 
\begin{equation}
\label{eq:JJI}
\mathbb{J}^I_{l_1 m_1 l_2 m_2} \left( fD_a, fD_b, \hat{k} \right) = J_{l_1 m_1}^R\left( fD_a, \hat{k} \right) J_{l_2 m_2}^{R*}\left( fD_b, \hat{k} \right) + J_{l_1 m_1}^L\left( fD_a, \hat{k} \right) J_{l_2 m_2}^{L*}\left( fD_b, \hat{k} \right) \,,
\end{equation}
\begin{equation}
\label{eq:JJV}
\mathbb{J}^V_{l_1 m_1 l_2 m_2} \left( fD_a, fD_b, \hat{k} \right) = J_{l_1 m_1}^R\left( fD_a, \hat{k} \right) J_{l_2 m_2}^{R*}\left( fD_b, \hat{k} \right) - J_{l_1 m_1}^L\left( fD_a, \hat{k} \right) J_{l_2 m_2}^{L*}\left( fD_b, \hat{k} \right) \,,
\end{equation}
\begin{equation}
\label{eq:JJQpU}
\mathbb{J}^{Q + i U}_{l_1 m_1 l_2 m_2} \left( fD_a, fD_b, \hat{k} \right) = J_{l_1 m_1}^L\left( fD_a, \hat{k} \right) J_{l_2 m_2}^{R*}\left( fD_b, \hat{k} \right) \,,
\end{equation}
and
\begin{equation}
\label{eq:JJQmU}
\mathbb{J}^{Q - i U}_{l_1 m_1 l_2 m_2} \left( fD_a, fD_b, \hat{k} \right) = J_{l_1 m_1}^R\left( fD_a, \hat{k} \right) J_{l_2 m_2}^{L*}\left( fD_b, \hat{k} \right) \,.
\end{equation}
For an isotropic unpolarized SGWB, $I \neq 0$ and $Q = U = V = 0$, hence $\gamma_{lm}^I \neq 0$ and $\gamma_{lm}^{V/Q/U} = 0$. Thus, a nontrivial $\gamma_{lm}^{V/Q/U}$ can be traced back to $V/Q/U \neq 0$. Information on the anisotropy and polarization of the SGWB are contained in the Stokes parameters \cite{AnilKumar:2023yfw, Liu:2022skj}.

The rest of this paper dwells on the derivation and analysis of the correlation for an anisotropic polarized SGWB with tensor, vector, and scalar gravitational d.o.f.s.

\section{Correlation for an anisotropic polarized SGWB: power spectrum formalism}
\label{sec:correlationforanisotropies}

In this section, we outline the general expressions pertaining to the correlations in the SGWB. The calculation of the $J_{lm}\left(fD , \hat{k}\right)$'s at the core of this section proceeds as in \cite{Bernardo:2022rif} and is reviewed in detail in Appendix \ref{sec:Jlmcalculations}.

\subsection{Tensor GW modes}
\label{subsec:tensorgwmodes}

For the tensor GWs, the $J_{lm}\left(fD , \hat{k}\right)$'s turn out to be
\begin{equation}
\label{eq:Jlm_tensor}
    J_{lm}^{R,L}\left(fD, \hat{k}\right) = - _{\mp 2}Y_{lm}^* \left( \hat{k} \right) e^{\mp 2i \alpha} \sqrt{2}\pi i^l \sqrt{ \dfrac{(l + 2)!}{(l - 2)!} } \int_0^{2\pi fDv} \dfrac{dx}{v} \ e^{ix/v} \dfrac{j_l(x)}{x^2} \,,
\end{equation}
where the upper (lower) signs correspond $R$ ($L$) polarizations, respectively. The factor $e^{i m' \alpha}$ is a redundant phase due to an extra rotational d.o.f. and does not enter the correlation.

In the infinite distance limit, $D \rightarrow \infty$, the integral admits the analytical expression
\begin{equation}
    \int_0^{\infty} \dfrac{dx}{v} \ e^{ix/v} \dfrac{j_l(x)}{x^2} = i \sqrt{\pi } 2^{-(l+1)} (i v)^{l-2} \Gamma (l-1) \, _2\tilde{F}_1\left(\frac{l-1}{2},\frac{l}{2};l+\frac{3}{2};v^2\right) 
\end{equation}
where $_2\tilde{F}_1\left( a, b; c; x \right) = _2 F_1\left( a, b; c; x \right)/\Gamma(c)$ is a regularized hypergeometric function. If $v = 1$, this reduces further to
\begin{equation}
    \int_0^\infty dx \ e^{ix} \dfrac{j_l(x)}{x^2} = 2 i^{l-1} \dfrac{(l - 2)!}{(l + 2)!} \,,
\end{equation}
which gives the HD correlation for an isotropic SGWB.

Moving forward, we expand the Stokes parameters for tensor modes in spin-weighted spherical harmonics:
\begin{eqnarray}
I\left(f, \hat{k}\right) &=& \sum_{lm} I_{lm}(f) Y_{lm}\left(\hat{k}\right) \,, \\
V\left(f, \hat{k}\right) &=& \sum_{lm} V_{lm}(f) Y_{lm}\left(\hat{k}\right) \,, \\
\left( Q + i U\right)\left(f,\hat{k}\right) &=& \left( Q + iU \right)_{lm}(f) \,_{+4}Y_{lm}\left(\hat{k}\right) \,, \\
\left( Q - i U\right)\left(f,\hat{k}\right) &=& \left( Q - iU \right)_{lm}(f) \,_{-4}Y_{lm}\left(\hat{k}\right) \,.
\end{eqnarray}
The above expression lets us write down
\begin{equation}
\begin{split}
    \mathbb{J}_{l_1 m_1 l_2 m_2}^{I,V} \left( fD_a, fD_b, \hat{k} \right) = (-1)^{m_1} \bigg[ & _2 Y_{l_1-m_1}\left(\hat{k}\right) \,_{-2}Y_{l_2 m_2}\left(\hat{k}\right) \\
    & \pm _{-2}Y_{l_1 -m_1}\left(\hat{k}\right) \,_2 Y_{l_2 m_2}\left(\hat{k}\right) \bigg] J^{\rm \bf T}_{l_1}\left( fD_a \right) J_{l_2}^{{\rm \bf T}*} \left( fD_b \right)
\end{split}
\end{equation}
and 
\begin{equation}
    \mathbb{J}_{l_1 m_1 l_2 m_2}^{Q \pm i U} \left( fD_a, fD_b, \hat{k} \right) = (-1)^{m_1} \,_{\mp 2} Y_{l_1 -m_1}\left( \hat{k} \right) \,_{\mp 2} Y_{l_2 m_2}\left( \hat{k} \right) e^{\pm 4 i \alpha} J^{\rm \bf T}_{l_1}\left( fD_a \right) J_{l_2}^{{\rm \bf T}*}\left( f D_b \right) 
\end{equation}
where $J^{\rm \bf T}_l(fD)$'s are given by
\begin{equation}
    J^{\rm \bf T}_l\left(fD\right) = \sqrt{2} \pi i^l \sqrt{\dfrac{(l + 2)!}{(l - 2)!}} \int_0^{2\pi fDv} \dfrac{dx}{v} \ e^{ix/v} \dfrac{j_l(x)}{x^2} \,.
\end{equation}
The correlation is therefore given by
\begin{equation}
\label{eq:gammaIV_tensor}
\begin{split}
    \gamma_{lm}^{I,V}\left( f D_a, fD_b; \hat{e}_a, \hat{e}_b \right) = & \sum_{l_1 m_1 l_2 m_2}
    (-1)^{m_1} \left[ 1 \pm (-1)^{l + l_1 + l_2} \right] \\
    & \ \ \times J^{\rm \bf T}_{l_1}\left( fD_a \right) J_{l_2}^{{\rm \bf T}*}\left( f D_b \right) Y_{l_1 m_1} \left( \hat{e}_a \right) Y_{l_2 m_2}^*\left( \hat{e}_b \right) \phantom{\dfrac{1}{1}} \\
    & \ \ \times \sqrt{ \dfrac{(2l+1)(2l_1 + 1)(2l_2 + 1)}{4\pi} }
    \left( \begin{array}{ccc}
    l & l_1 & l_2 \\
    0 & -2 & 2
    \end{array} \right)
    \left( \begin{array}{ccc}
    l & l_1 & l_2 \\
    m & -m_1 & m_2
    \end{array} \right)
\end{split}
\end{equation}
and
\begin{equation}
\label{eq:gammaQU_tensor}
\begin{split}
    \gamma_{lm}^{Q \pm i U}\left( f D_a, fD_b; \hat{e}_a, \hat{e}_b \right) = & \sum_{l_1 m_1 l_2 m_2}
    (-1)^{m_1} J^{\rm \bf T}_{l_1}\left( fD_a \right) J_{l_2}^{{\rm \bf T}*}\left( f D_b \right) Y_{l_1 m_1} \left( \hat{e}_a \right) Y_{l_2 m_2}^*\left( \hat{e}_b \right) \\
    & \ \ \times \sqrt{ \dfrac{(2l+1)(2l_1 + 1)(2l_2 + 1)}{4\pi} }
    \left( \begin{array}{ccc}
    l & l_1 & l_2 \\
    \mp 4 & \pm 2 & \pm 2
    \end{array} \right)
    \left( \begin{array}{ccc}
    l & l_1 & l_2 \\
    m & -m_1 & m_2
    \end{array} \right) \,.
\end{split}
\end{equation}
Once again, $\gamma_{lm}^I \neq 0$ and $\gamma_{lm}^{V/Q/U} = 0$ correspond to the isotropic case. The above formulae thus give the most general expression for the correlation due to an anisotropic polarized SGWB with subluminal tensor GWs and finite pulsar distances.

The isotropic SGWB is recovered from this as the $l = m = 0$ special limit. This is easy to show using the identity
\begin{equation}
    \left( \begin{array}{ccc}
    0 & l_1 & l_2 \\
    0 & -2 & 2
    \end{array} \right)
    \left( \begin{array}{ccc}
    0 & l_1 & l_2 \\
    0 & -m_1 & m_2
    \end{array} \right) = \dfrac{(-1)^{m_1}}{2 l_1 + 1} \delta_{l_1 l_2} \delta_{m_1 m_2} \ \ \ \ \ , l_1 \geq 2, m_1 = -l_1, -l_1 + 1, \cdots l_1 - 1, l_1 \,,
\end{equation}
and the addition theorem
\begin{equation}
\label{eq:addition_theorem}
    P_l \left( \hat{e}_a \cdot \hat{e}_b \right) = \dfrac{4\pi}{2l+1} \sum_{m} Y_{lm}\left(\hat{e}_a\right) Y_{lm}^*\left(\hat{e}_b\right) \,.
\end{equation}
It follows that
\begin{equation}
    \gamma_{00}^V = \gamma_{00}^{Q \pm i U} = 0
\end{equation}
and
\begin{equation}
    \gamma^I_{00} = \sum_l \dfrac{2l + 1}{4\pi} C_l P_l \left(\hat{e}_a \cdot \hat{e}_b \right)
\end{equation}
where the tensor GW power spectrum is given by \cite{Bernardo:2022rif}
\begin{equation}
    C_l^{\rm \bf T} = \dfrac{J^{\rm \bf T}_l\left(f D_a\right) J^{{\rm \bf T}*}_l\left(f D_b\right)}{\sqrt{\pi}} \,.
\end{equation}
The HD correlation power spectrum emerges from this in the luminal GW ($v = 1$) and infinite pulsar distance ($fD_{a,b} \gg 1$) limit:
\begin{equation}
    C_l^{\rm \bf T} \sim \dfrac{8\pi^{3/2}}{(l + 2)(l + 1)l(l - 1)} \,.
\end{equation}

We close this part by writing the correlations in the computational frame such that pulsar $a$ is on the $\hat{z}$ axis and pulsar $b$ is on the $x-z$ plane with $\hat{e}_a \cdot \hat{e}_b = \cos \zeta$. Using spherical harmonic identities,
\begin{equation}
\label{eq:Yea_zz}
    Y_{l_1 m_1}\left( \hat{e}_a \right) = \sqrt{ \dfrac{2l_1 + 1}{4\pi} } \delta_{m_1 0}
\end{equation}
and
\begin{equation}
\label{eq:Yeb_xz}
    Y_{l_2 m_2}^*\left( \hat{e}_b \right) = Y_{l_2 m_2}^*\left( \zeta, 0 \right) = Y_{l_2 m_2}\left( \zeta, 0 \right) \,,
\end{equation}
we obtain the useful expressions
\begin{equation}
\label{eq:gammaIV_tensor_cframe}
\begin{split}
    \gamma_{lm}^{I,V}\left( f D_a, fD_b, \zeta \right) = \sum_{l_1 l_2}
    & (-1)^m \left( \dfrac{2 l_1 + 1}{4\pi} \right) \left[ 1 \pm (-1)^{l + l_1 + l_2} \right] J^{\rm \bf T}_{l_1}\left( fD_a \right) J_{l_2}^{{\rm \bf T}*}\left( f D_b \right) Y_{l_2 m}\left( \zeta, 0 \right) \\
    & \sqrt{(2l+1)(2l_2 + 1)}
    \left( \begin{array}{ccc}
    l & l_1 & l_2 \\
    0 & -2 & 2
    \end{array} \right)
    \left( \begin{array}{ccc}
    l & l_1 & l_2 \\
    m & 0 & -m
    \end{array} \right)
\end{split}
\end{equation}
and
\begin{equation}
\label{eq:gammaQU_tensor_cframe}
\begin{split}
    \gamma_{lm}^{Q \pm i U}\left( f D_a, fD_b, \zeta \right) = \sum_{l_1 l_2}
    & (-1)^m \left( \dfrac{2l_1+1}{4\pi} \right) J^{\rm \bf T}_{l_1}\left( fD_a \right) J_{l_2}^{{\rm \bf T}*}\left( f D_b \right) Y_{l_2 m}\left( \zeta, 0 \right) \\
    & \sqrt{ (2l+1)(2l_2 + 1) }
    \left( \begin{array}{ccc}
    l & l_1 & l_2 \\
    \mp 4 & \pm 2 & \pm 2
    \end{array} \right)
    \left( \begin{array}{ccc}
    l & l_1 & l_2 \\
    m & 0 & -m
    \end{array} \right) \,.
\end{split}
\end{equation}
Owing to the 3j symbol, it should be noted that the correlation components satisfy the following parity relations:
\begin{eqnarray}
    \gamma^{I}_{l-m} &=& (-1)^{m} \gamma^I_{lm} \,, \label{eq:ParitymTensorI} \\
    \gamma^{V}_{l-m} &=& (-1)^{m + 1} \gamma^V_{lm} \,, \label{eq:ParitymTensorV} \\
    \gamma^{Q \pm i U}_{l-m} &=& (-1)^{m} \gamma^{Q \mp i U}_{lm} \label{eq:ParitymTensorQiU} \,.
\end{eqnarray}
This can be used to obtain the tensor SGWB correlations $\gamma_{l-|m|}$ provided $\gamma_{lm}$ with $m > 0$.

\subsection{Vector GW modes}
\label{subsec:vectorgwmodes}

For vector GWs, the relevant $J_{lm}\left(fD, \hat{k}\right)$'s are given by
\begin{equation}
\label{eq:Jlm_vector}
\begin{split}
    J_{(l \geq 1)m}^{VR,VL} \left( fD, \hat{k} \right) = \mp \, _{\mp 1} Y_{lm}^*\left(\hat{k}\right) e^{\mp i \alpha} 2\sqrt{2} \pi i^l \sqrt{l (l + 1)} \int_0^{2\pi f D v} \dfrac{dx}{v} \ e^{ix/v} \dfrac{d}{dx} \left( \dfrac{j_l(x)}{x} \right) \,.
\end{split}
\end{equation}
The phase factor $e^{\pm i \alpha}$ corresponds to an arbitrary rotational d.o.f. that drops out in the correlation. On the one hand, the integral is conveniently expressed as
\begin{equation}
\label{eq:Jlm_vec}
\begin{split}
    \int_0^{2\pi f D v} \dfrac{dx}{v} \ e^{ix/v} \dfrac{d}{dx} \left( \dfrac{j_l(x)}{x} \right) = & - \dfrac{i}{v} \int_0^{2\pi f D v} \dfrac{dx}{v} \ e^{ix/v} \dfrac{j_l(x)}{x} \\ 
    & \ \ + \dfrac{e^{2\pi i fD}}{v} \dfrac{j_l\left(2\pi f D v\right)}{2\pi f D v} - \dfrac{\sqrt{\pi } 2^{-(l + 1)}}{v \Gamma \left(l+(3/2)\right)} \epsilon^{l - 1}|_{\epsilon \rightarrow 0^+} \,.
\end{split}
\end{equation}
 The second boundary term is associated with finite pulsar distances; the third boundary term reduces to a constant ($\sim 1/v$) for $l = 1$ (dipole moment) but vanishes for all $l > 1$.

The integral admits an analytical expression for the infinite distance case,
\begin{equation}
    \int_0^{\infty} \dfrac{dx}{v} \ e^{ix/v} \dfrac{j_l(x)}{x} = \sqrt{\pi } 2^{-(l+1)} i^l v^{l - 1} \Gamma (l) \, _2\tilde{F}_1\left(\frac{l}{2},\frac{l+1}{2};l+\frac{3}{2};v^2\right) \,,
\end{equation}
and further in the luminal GW limit,
\begin{equation}
    \int_0^{\infty} dx \ e^{ix} \dfrac{j_l(x)}{x} = i^l \dfrac{(l - 1)!}{(l + 1)!} \,.
\end{equation}
This leads to a correlation analogous to the HD correlation ($v = 1$ and $fD \rightarrow \infty$) for vector GWs.

As with the tensor modes, the step ahead demands us to expand the Stokes parameters, this time for vector modes, in spin-weighted spherical harmonics:
\begin{eqnarray}
I\left(f, \hat{k}\right) &=& \sum_{lm} I_{lm}(f) Y_{lm}\left(\hat{k}\right) \,, \\
V\left(f, \hat{k}\right) &=& \sum_{lm} V_{lm}(f) Y_{lm}\left(\hat{k}\right) \,, \\
\left( Q + i U\right)\left(f,\hat{k}\right) &=& \left( Q + iU \right)_{lm}(f) \,_{+2}Y_{lm}\left(\hat{k}\right) \,, \\
\left( Q - i U\right)\left(f,\hat{k}\right) &=& \left( Q - iU \right)_{lm}(f) \,_{-2}Y_{lm}\left(\hat{k}\right) \,.
\end{eqnarray}
We now derive the correlation for an anisotropic polarized SGWB with vector GW d.o.f.s. Following the formulae (\ref{eq:JJI}-\ref{eq:Jlm_tensor}), we obtain
\begin{equation}
\begin{split}
    \mathbb{J}_{l_1 m_1 l_2 m_2}^{I,V} \left( fD_a, fD_b, \hat{k} \right) = (-1)^{m_1 + 1} \bigg[ & _1 Y_{l_1-m_1}\left(\hat{k}\right) \,_{-1}Y_{l_2 m_2}\left(\hat{k}\right) \\
    & \pm _{-1}Y_{l_1 -m_1}\left(\hat{k}\right) \,_1 Y_{l_2 m_2}\left(\hat{k}\right) \bigg] J^{\rm \bf V}_{l_1}\left( fD_a \right) J_{l_2}^{{\rm \bf V}*} \left( fD_b \right)
\end{split}
\end{equation}
and
\begin{equation}
    \mathbb{J}_{l_1 m_1 l_2 m_2}^{Q \pm i U} \left( fD_a, fD_b, \hat{k} \right) = (-1)^{m_1} \,_{\mp 1} Y_{l_1 -m_1}\left( \hat{k} \right) \,_{\mp 1} Y_{l_2 m_2}\left( \hat{k} \right) e^{\pm 2 i \alpha} J^{\rm \bf V}_{l_1}\left( fD_a \right) J_{l_2}^{{\rm \bf V}*}\left( f D_b \right) 
\end{equation}
where the $J^{\rm \bf V}_l(fD)$'s are given by
\begin{equation}
    J^{\rm \bf V}_l\left(fD\right) = 2 \sqrt{2} \pi i^l \sqrt{l(l+1)} \int_0^{2\pi fDv} \dfrac{dx}{v} \ e^{ix/v} \dfrac{d}{dx} \left( \dfrac{j_l(x)}{x} \right) \,.
\end{equation}
The correlation is therefore obtained by the sums
\begin{equation}
\label{eq:gammaIV_vector}
\begin{split}
    \gamma_{lm}^{I,V}\left( f D_a, fD_b; \hat{e}_a, \hat{e}_b \right) = & \sum_{l_1 m_1 l_2 m_2}
    (-1)^{m_1 + 1} \left[ 1 \pm (-1)^{l + l_1 + l_2} \right] \\
    & \ \ \times J^{\rm \bf V}_{l_1}\left( fD_a \right) J_{l_2}^{{\rm \bf V}*}\left( f D_b \right) Y_{l_1 m_1} \left( \hat{e}_a \right) Y_{l_2 m_2}^*\left( \hat{e}_b \right) \phantom{\dfrac{1}{1}} \\
    & \ \ \times \sqrt{ \dfrac{(2l+1)(2l_1 + 1)(2l_2 + 1)}{4\pi} }
    \left( \begin{array}{ccc}
    l & l_1 & l_2 \\
    0 & -1 & 1
    \end{array} \right)
    \left( \begin{array}{ccc}
    l & l_1 & l_2 \\
    m & -m_1 & m_2
    \end{array} \right)
\end{split}
\end{equation}
and
\begin{equation}
\label{eq:gammaQU_vector}
\begin{split}
    \gamma_{lm}^{Q \pm i U}\left( f D_a, fD_b; \hat{e}_a, \hat{e}_b \right) = & \sum_{l_1 m_1 l_2 m_2}
    (-1)^{m_1} J^{\rm \bf V}_{l_1}\left( fD_a \right) J_{l_2}^{{\rm \bf V}*}\left( f D_b \right) Y_{l_1 m_1} \left( \hat{e}_a \right) Y_{l_2 m_2}^*\left( \hat{e}_b \right) \\
    & \ \ \times \sqrt{ \dfrac{(2l+1)(2l_1 + 1)(2l_2 + 1)}{4\pi} }
    \left( \begin{array}{ccc}
    l & l_1 & l_2 \\
    \mp 2 & \pm 1 & \pm 1
    \end{array} \right)
    \left( \begin{array}{ccc}
    l & l_1 & l_2 \\
    m & -m_1 & m_2
    \end{array} \right) \,.
\end{split}
\end{equation}

The isotropic case is simply obtained by noting that
\begin{equation}
    \left( \begin{array}{ccc}
    0 & l_1 & l_2 \\
    0 & -1 & 1
    \end{array} \right)
    \left( \begin{array}{ccc}
    0 & l_1 & l_2 \\
    0 & -m_1 & m_2
    \end{array} \right) = \dfrac{(-1)^{m_1 + 1}}{2 l_1 + 1} \delta_{l_1 l_2} \delta_{m_1 m_2} \ \ \ \ \ , l_1 \geq 1, m_1 = -l_1, -l_1 + 1, \cdots l_1 - 1, l_1 \,.
\end{equation}
Then, it follows that
\begin{equation}
    \gamma_{00}^V = \gamma_{00}^{Q \pm i U} = 0
\end{equation}
and
\begin{equation}
    \gamma^I_{00} = \sum_l \dfrac{2l + 1}{4\pi} C_l P_l \left(\hat{e}_a \cdot \hat{e}_b \right)
\end{equation}
where the vector GW correlation power spectrum is given by \cite{Bernardo:2022rif}
\begin{equation}
    C_l^{\rm \bf V} = \dfrac{J^{\rm \bf V}_l\left(f D_a\right) J^{{\rm \bf V}*}_l\left(f D_b\right)}{\sqrt{\pi}} \,.
\end{equation}

As with tensor GWs, we leave the section by giving the expression for the correlations in the computational frame. Using \eqref{eq:Yea_zz} and \eqref{eq:Yeb_xz}, we end up with the useful expressions,
\begin{equation}
\label{eq:gammaIV_vector_cframe}
\begin{split}
    \gamma_{lm}^{I,V}\left( f D_a, fD_b, \zeta \right) = \sum_{l_1 l_2}
    & (-1)^{1 + m} \left( \dfrac{2l_1 + 1}{4\pi} \right) \left[ 1 \pm (-1)^{l + l_1 + l_2} \right] J^{\rm \bf V}_{l_1}\left( fD_a \right) J_{l_2}^{{\rm \bf V}*}\left( f D_b \right) Y_{l_2 m}\left( \zeta, 0 \right) \\
    & \sqrt{ (2l+1)(2l_2 + 1) }
    \left( \begin{array}{ccc}
    l & l_1 & l_2 \\
    0 & -1 & 1
    \end{array} \right)
    \left( \begin{array}{ccc}
    l & l_1 & l_2 \\
    m & 0 & -m
    \end{array} \right)
\end{split}
\end{equation}
and
\begin{equation}
\label{eq:gammaQU_vector_cframe}
\begin{split}
    \gamma_{lm}^{Q \pm i U}\left( f D_a, fD_b, \zeta \right) = \sum_{l_1 l_2}
    & (-1)^m \left( \dfrac{2l_1 + 1}{4\pi} \right) J^{\rm \bf V}_{l_1}\left( fD_a \right) J_{l_2}^{{\rm \bf V}*}\left( f D_b \right) Y_{l_2 m}\left( \zeta, 0 \right)  \\
    & \sqrt{ (2l+1)(2l_2 + 1) }
    \left( \begin{array}{ccc}
    l & l_1 & l_2 \\
    \mp 2 & \pm 1 & \pm 1
    \end{array} \right)
    \left( \begin{array}{ccc}
    l & l_1 & l_2 \\
    m & 0 & -m
    \end{array} \right) \,.
\end{split}
\end{equation}
The correlations $\gamma_{l-|m|}$ can be recovered with $\gamma_{lm}$ for $m>0$ using the same relations as with the tensor case (\ref{eq:ParitymTensorI}-\ref{eq:ParitymTensorQiU}).

\subsection{Scalar GW modes}
\label{subsec:scalargwmodes}

The $J_{lm}\left(fD, \hat{k}\right)$'s for scalar GWs simplify to
\begin{equation}
\label{eq:JlmSTfinal}
\begin{split}
    J^\text{ST}_{lm}\left( fD, \hat{k} \right) =  Y^*_{lm}\left(\hat{k}\right) \left( 2\pi i^l \right) \int_0^{2\pi fDv} \dfrac{d x}{v} \ e^{i x/v} \left( j_l''(x) + j_l(x) \right)  \,
\end{split}
\end{equation}
and
\begin{equation}
\label{eq:JlmSLfinal}
\begin{split}
    \dfrac{J^\text{SL}_{lm}\left( fD, \hat{k} \right)}{\sqrt{2}} =  - Y^*_{lm}\left(\hat{k}\right) \left( 2\pi i^l \right) \int_0^{2\pi fDv} \dfrac{d x}{v} \ e^{i x/v} j_l''\left(x\right) \,,
\end{split}
\end{equation}
respectively, for the ST and SL GW polarizations. It is worth noting that the ST integral can be recast as a total boundary for $v = 1$, i.e., note that
\begin{equation}
\label{eq:JS_boundary}
    e^{ix/v} \left( j_l''(x) + j_l(x) \right) = \dfrac{d}{dx} \left[ e^{ix/v} \left( j_l'(x) - \dfrac{i}{v} j_l(x) \right) \right] + \dfrac{v^2 - 1}{v^2} e^{ix/v} j_l\left(x\right) \,.
\end{equation}
Furthermore, given the asymptotic expansion
\begin{equation}
    e^{ix/v} \left( j_l'(x) - \dfrac{i}{v} j_l(x) \right) \sim \dfrac{\sqrt{\pi } 2^{-(l+1)}}{\Gamma \left(l+(3/2)\right)} x^l \left( \dfrac{l}{x} + \dfrac{i}{v}(l - 1) + O\left( x \right) \right) \ \ , \ \ \ \ x \rightarrow 0^+
\end{equation}
and the integral identity
\begin{equation}
    \int_0^{r} d x \ e^{i x} j_l\left(x\right) = 2^l r^{l+1} \Gamma (l+1)^2 \, _2\tilde{F}_2(l+1,l+1;l+2,2 l+2;2 i r) \ \ , \ \ \ \ \text{Re}(l) > -1 \,, 
\end{equation}
we find that the ST and SL integrals take analytical closed forms for finite $fD$ and $v = 1$:
\begin{equation}
\int_0^{2\pi fD} d x \ e^{i x} \left( j_l''(x) + j_l(x) \right) = e^{2\pi i fD} \left[ j_l'\left(2\pi fD\right) - i j_l\left(2\pi fD\right) \right]
\end{equation}
and
\begin{equation}
\begin{split}
\int_0^{2\pi fD} d x \ e^{i x} j_l''(x) = \, & e^{2\pi i fD} \left[ j_l'\left(2\pi fD\right) - i j_l\left(2\pi fD\right) \right] \\
& \ \ - 2^l \left(2\pi fD\right)^{l+1} \Gamma (l+1)^2 \, _2\tilde{F}_2 \left( l+1,l+1;l+2,2 l+2;4\pi i fD \right) \,.
\end{split}
\end{equation}
This shows that the ST terms reduce to a dipolar contribution; on the other hand, the SL modes are characterized by a diverging higher moment contributions in the infinite distance limit. Utilizing \eqref{eq:JS_boundary}, we also obtain an analytical expression for infinite pulsar distances and subluminal scalar GWs:
\begin{equation}
    \int_0^\infty \dfrac{dx}{v} \ e^{ix/v} j_l(x) = \sqrt{\pi } 2^{-(l+1)} (i v)^{l+1} \Gamma (l+1) \, _2\tilde{F}_1\left(\frac{l+1}{2},\frac{l+2}{2};l+\frac{3}{2};v^2\right) \,.
\end{equation}

A takeaway is that the $J_{lm}\left(f D, \hat{k}\right)$'s for the scalar GWs can be written as
\begin{equation}
    J_{lm}\left(fD, \hat{k}\right) = Y^*_{lm}\left(\hat{k}\right) \mathcal{F}^{\rm \bf S}_l\left(f D\right) \,,
\end{equation}
where the $\mathcal{F}^{\rm \bf S}_l\left(fD\right)$'s can be read from \eqref{eq:JlmSTfinal} and \eqref{eq:JlmSLfinal}. The angular and radial dependencies separate. In this case, we calculate the correlation of an unpolarized SGWB directly with \eqref{eq:orf_general}. For either ST or SL modes, we obtain
\begin{equation}
\label{eq:gamma_scalar}
\begin{split}
    \gamma_{lm} \left( fD_a, fD_b; \hat{e}_a, \hat{e}_b \right) = \sum_{l_1 m_1 l_2 m_2}
    & \mathcal{F}^{\rm \bf S}_{l_1}\left( fD_a \right) \mathcal{F}_{l_2}^{{\rm \bf S} *}\left( f D_b \right) Y_{l_1 m_1} \left( \hat{e}_a \right) Y_{l_2 m_2}^*\left( \hat{e}_b \right) \\
    & \sqrt{ \dfrac{(2l+1)(2l_1 + 1)(2l_2 + 1)}{4\pi} }
    \left( \begin{array}{ccc}
    l & l_1 & l_2 \\
    0 & 0 & 0
    \end{array} \right)
    \left( \begin{array}{ccc}
    l & l_1 & l_2 \\
    m & -m_1 & m_2
    \end{array} \right) \,.
\end{split}
\end{equation}

To recover the isotropic case, $l = m = 0$, we use the identity
\begin{equation}
    \left( \begin{array}{ccc}
    0 & l_1 & l_2 \\
    0 & 0 & 0
    \end{array} \right)
    \left( \begin{array}{ccc}
    0 & l_1 & l_2 \\
    0 & -m_1 & m_2
    \end{array} \right) = \dfrac{(-1)^{m_1}}{2 l_1 + 1} \delta_{l_1 l_2} \delta_{m_1 m_2} \ \ \ \ \ , l_1 \geq 0, m_1 = -l_1, -l_1 + 1, \cdots l_1 - 1, l_1 \,,
\end{equation}
and the addition theorem \eqref{eq:addition_theorem}. This leads to the correlation
\begin{equation}
    \gamma_{00} \left( fD_a, fD_b; \hat{e}_a, \hat{e}_b \right) = \sum_l \dfrac{2l+1}{4\pi} C_l P_l\left( \hat{e}_a \cdot \hat{e}_b \right) \,,
\end{equation}
where the scalar GW correlation power spectrum is given by
\begin{equation}
    C_l^{\rm \bf S} = \dfrac{\mathcal{F}^{\rm \bf S}_l\left(f D_a \right) \mathcal{F}_l^{{\rm \bf S} *}\left( fD_b\right)}{\sqrt{4\pi}} \, .
\end{equation}
We make the further identification
\begin{equation}
    \mathcal{F}^{\rm \bf S}_l(fD) = \sqrt{32 \pi^2} i^{l - 1} F_l^{\rm \bf S}\left(fD\right)
\end{equation}
where
\begin{equation}
    F_l^{\rm \bf S}(fD) = - \dfrac{i}{2} \int_0^{2\pi fDv} \dfrac{dx}{v} \ e^{ix/v} R^{\rm \bf S}_l\left(x\right) \,.
\end{equation}
Above, $R_l^{\rm \bf SL}(x) = j_l''(x)$ for the SL case and $R_l^{\rm \bf ST}(x) = -\left( R_l^{\rm \bf SL}(x) + j_l(x) \right)/\sqrt{2}$ for the ST case  \cite{Qin:2020hfy}. The $F_l(\infty)$'s are the projection factors considered in \cite{Qin:2020hfy} such that $C_l \propto 32 \pi^2 | F_l(\infty) |^2$. These are generalized to finite distances by the above expressions.

An interesting result emerging from generalized scalar-tensor theory is that the ST and SL modes interfere to leave only dominant monopolar and dipolar traces in the correlation \cite{Qin:2020hfy, Bernardo:2022vlj, Bernardo:2023pwt}; in symbols,
\begin{equation}
    h_{AB} \propto \varepsilon_{AB}^{\text{ST}} + \varepsilon_{AB}^{\text{SL}} \left(1-v^2\right)/\sqrt{2} \,,
\end{equation}
where $v$ is the group speed of scalar GWs. This paper extends this result to an anisotropic polarized SGWB. The particular superposition translates to  an overall relativistic scalar-induced projection factor,
\begin{equation}
    {F}^{\rm \bf S}_l \left(fD\right) = - \dfrac{i}{2} \int_0^{2\pi fDv} \dfrac{dx}{v} e^{ix/v} \left( R^{\rm \bf ST}_l(x) + \dfrac{1-v^2}{\sqrt{2}} R^{\rm \bf SL}_l(x)  \right) \,,
\end{equation}
that reduces to a total boundary. In the limit $D \rightarrow \infty$, all the $l \geq 2$ scalar modes vanish. In general, for finite $D \sim 0.1-1$ kpc relevant for PTA, the spectrum is characterized by a dominant monopole and a dipole. In this case, the higher $l \geq 2$ are nontrivial but remain subdominant. We shall use the overall scalar field-induced correlation in our analysis.

As with the tensor and vector GW cases, we close this section by expressing the correlation in the computational frame. Using the identities \eqref{eq:Yea_zz} and \eqref{eq:Yeb_xz}, we obtain
\begin{equation}
\label{eq:gamma_scalar_cframe}
\begin{split}
    \gamma_{lm} \left( fD_a, fD_b, \zeta \right) = \sum_{l_1 l_2}
    & (-1)^m \left( \dfrac{2l_1 + 1}{4\pi} \right) \mathcal{F}^{\rm \bf S}_{l_1}\left( fD_a \right) \mathcal{F}_{l_2}^{{\rm \bf S} *}\left( f D_b \right) Y_{l_2 m}\left( \zeta, 0 \right) \\
    & \sqrt{ (2l+1)(2l_2 + 1) }
    \left( \begin{array}{ccc}
    l & l_1 & l_2 \\
    0 & 0 & 0
    \end{array} \right)
    \left( \begin{array}{ccc}
    l & l_1 & l_2 \\
    m & 0 & -m
    \end{array} \right) \,.
\end{split}
\end{equation}
Note that for the scalar case, $l + l_1 + l_2$ must be an even integer, and so the correlations $\gamma_{l-|m|}$ can be evaluated simply using
\begin{equation}
    \gamma_{l-|m|} = (-1)^m \gamma_{lm} \,,
\end{equation}
where $m$ is a positive integer.

\section{Discussion}
\label{sec:discussion}

In this section, we present the correlation signals at various levels of anisotropy for tensor (Sec. \ref{subsec:tensorgwcorrelations}), vector (Sec. \ref{subsec:vectorgwcorrelations}), and scalar (Sec. \ref{subsec:scalargwcorrelations}) modes in the SGWB.

\subsection{Tensor GW correlations}
\label{subsec:tensorgwcorrelations}

The natural starting point of the discussion is the isotropic case ($l = m = 0$), for which case the HD curve ($v = 1$) is reproduced. We remind that from a gravitational physics perspective, this correlation pattern is produced by the $+$ and $\times$ GW polarizations, producing the dominantly quadrupolar signal expected in GR. We now discuss alternatives to the isotropic HD correlation.

\begin{figure}[h!]
\center
	\subfigure[ \ $l = 0$ ]{
		\includegraphics[width = 0.475 \textwidth]{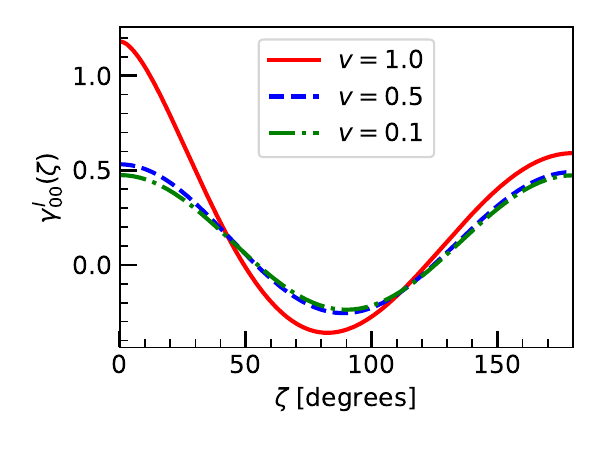}
		}
	\subfigure[ \ $l = 1$ ]{
		\includegraphics[width = 0.475 \textwidth]{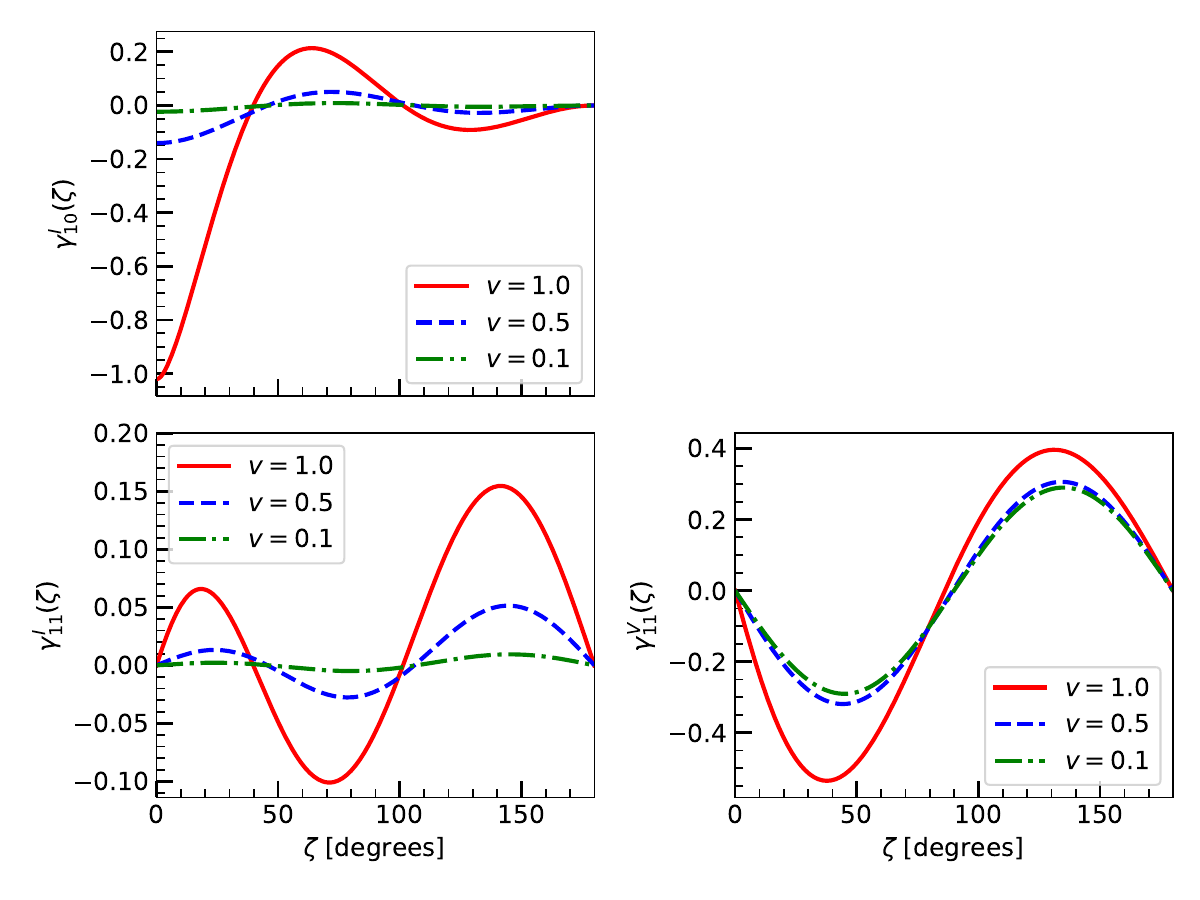}
		}
\caption{(a) Correlations induced by tensor modes for $l = m = 0$ (isotropic unpolarized case) and (b) for $l = 1$ for $v = 1.0, 0.5, 0.1$ with $fD = 10^3$.}
\label{fig:gammal01_T}
\end{figure}

For the tensor GWs, the HD curve is shown in Figure \ref{fig:gammal01_T} together with two other cases corresponding to GW speeds $v = 0.5$ and $0.1$. For the $v < 1$ cases, we recover the known enhancement of the quadrupolar power in the isotropic case \cite{Qin:2020hfy, Bernardo:2022rif}. For more discussion of the isotropic SGWB with nonluminal GW propagation, we refer to \cite{Qin:2018yhy, Qin:2020hfy, Bernardo:2022rif, Bernardo:2022xzl, Liang:2023ary}.

In this context, the HD curve appears a special case of the isotropic SGWB, i.e., $\gamma_{00}^{I} \sim \gamma^{\rm HD}(\zeta)$ with everything else vanishing, $\gamma_{lm}^{V} = \gamma_{lm}^{Q \pm i U} = 0$ for $l > 0$ and $|m| \leq l$. Thus, we look toward the more general and interesting scenario of an anisotropic polarized SGWB induced by gravitational tensor modes. It is useful to keep in mind that the correlation $\gamma_{lm}^V(\zeta)$ vanishes for $m = 0$, and the $\gamma_{lm}^{Q \pm iU}(\zeta)$ becomes nontrivial only for $l \geq 4$ owing to the Wigner-3j symbol, reflecting the spin-2 nature of tensor GWs.

\begin{figure}[h!]
\center
	\subfigure[ \ $l = 2$ ]{
		\includegraphics[width = 0.475 \textwidth]{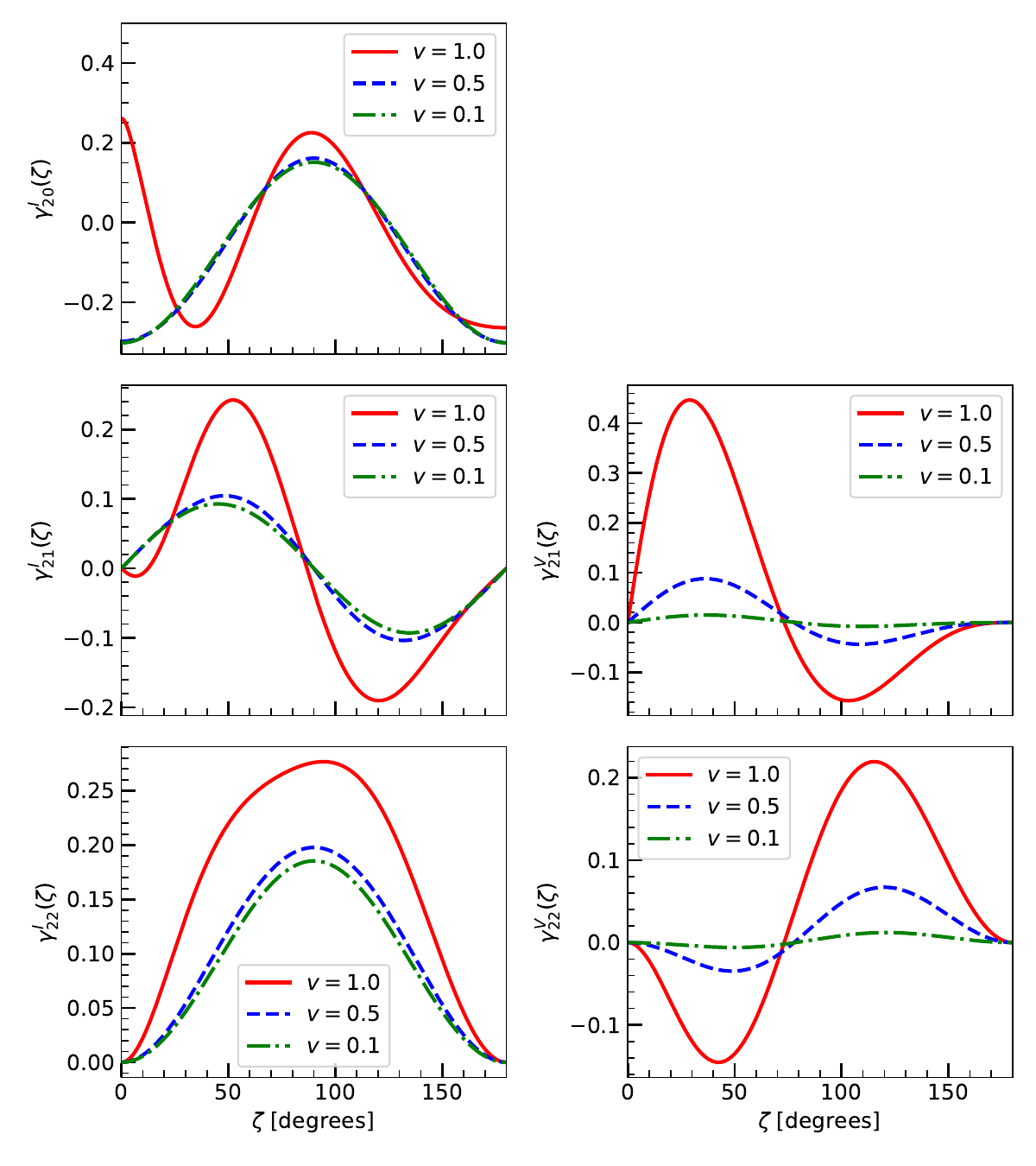}
		}
	\subfigure[ \ $l = 3$ ]{
		\includegraphics[width = 0.375 \textwidth]{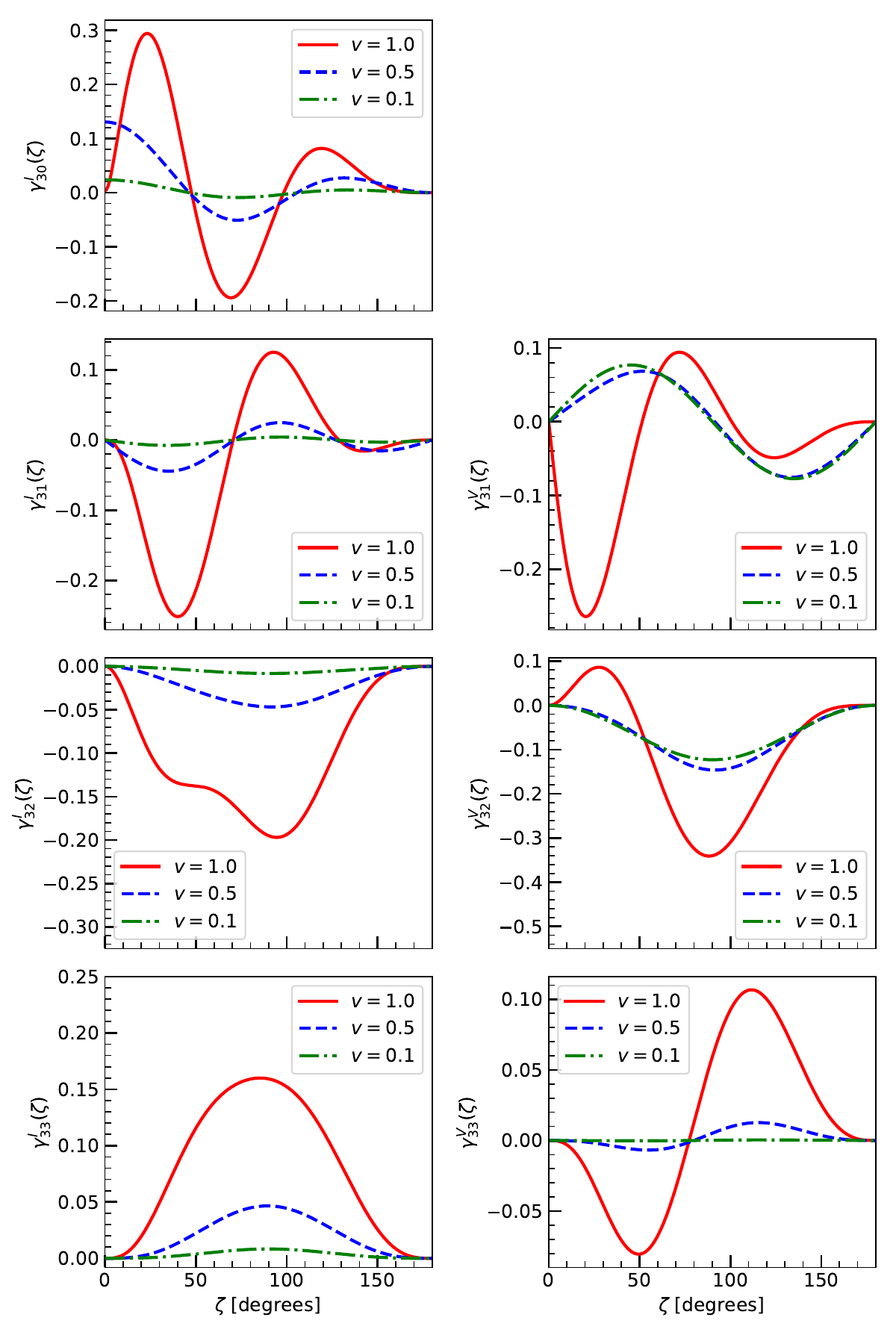}
		}
\caption{(a) Correlations induced by tensor modes for $l = 2$ and (b) for $l = 3$ for $v = 1.0, 0.5, 0.1$ with $fD = 10^3$.}
\label{fig:gammal23_T}
\end{figure}

\begin{figure}[h!]
    \centering
    \includegraphics[width = 0.95 \textwidth]{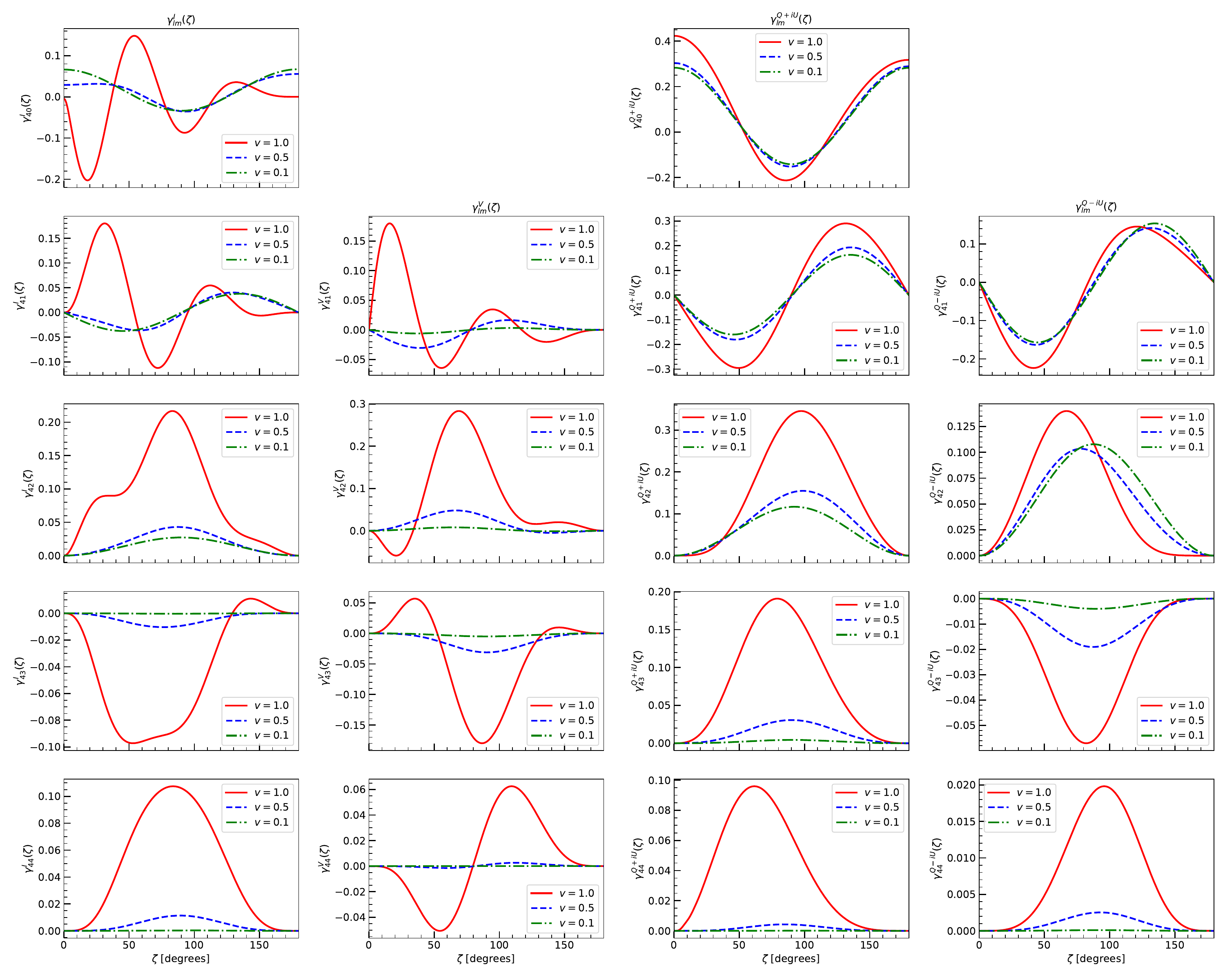}
    \caption{Correlations induced by tensor modes for $l = 4$ for $v = 1.0, 0.5, 0.1$ with $fD = 10^3$.}
    \label{fig:gammal4_T}
\end{figure}

The nontrivial modes and moments of the tensor correlation up to $l = 4$ are shown in Figures \ref{fig:gammal01_T}, \ref{fig:gammal23_T}, and \ref{fig:gammal4_T}. Notably, the previous discussion is reflected in this presentation such that for $1 \leq l \leq 3$ we show only the $I$ and $V$ contributions to the correlation since the $Q$ and $U$ components vanish. Then, at $l = 4$ we find nontrivial contributions to the correlation from all Stokes parameters, indicating anisotropy and polarization.

We find that subluminal GW propagation generally damps out the anisotropic and polarized components of the correlation. In the cases shown, the correlation with the maximum size always belongs to the luminal one. We may trace this back to the fact that in the isotropic case, the quadrupole dominates the correlation for subluminal travelling waves and we are seeing this manifest to the anisotropic and polarized components in Figures \ref{fig:gammal01_T}--\ref{fig:gammal4_T}. The drop in the power is associated to the declining contributions from the higher multipoles for subluminal GWs \cite{Qin:2020hfy, Bernardo:2022rif, Mihaylov:2019lft}. However, this should be taken no more than as a rule of thumb as there are clear cases when the correlations produced by subluminal GWs remain comparable in size to the luminal one. Notable modes displaying this behaviour are $\gamma_{11}^V$, $\gamma_{2m}^I$, $\gamma_{31}^V$, $\gamma_{32}^V$, $\gamma_{4m}^{I/Q \pm iU}$, with $m \leq 2$. Turning this around, one may look toward these interesting components of the correlation to look for signatures of an anisotropic polarized SGWB from subluminal GWs.

Another interesting place to look for differentiating between the luminal and subluminal GWs are the $\gamma^I_{l0}$ modes. This is utilizing the observation that its values and behavior at particularly small angular separations are notably different, e.g., $\gamma^I_{20} > 0$ for the luminal case but $\gamma^I_{20} < 0$ in the subluminal cases. A practical importance of searching for the higher-order, $l \geq 2$, anisotropy and polarization components for tensor GWs also arises: at higher modes, the correlations for different GW speeds do not manifest merely as an overall change in scale for different $v$'s but rather a totally different angular pattern is displayed. A basis of this is that the shape of the correlation is shared for most cases by the subluminal cases presented in Figures \ref{fig:gammal23_T}-\ref{fig:gammal4_T} whereas the luminal case paints an almost completely different angular correlation signature. We may look at the $\gamma_{30}^I$ and $\gamma_{31}^V$ components as a representative to elaborate on this point; for both of these, the correlation reaches an extremum for $\zeta \lesssim 50^\circ$ in the luminal case but not for the subluminal cases. There are many lookout options for this interesting behaviour at higher anisotropy and polarization modes. This is a compelling reason to consider them for establishing rigorous tests of gravity in the nanohertz regime.

\subsection{Vector GW correlations}
\label{subsec:vectorgwcorrelations}

As with tensor modes, the reasonable start of discussion of anisotropy and polarization in the correlation is the isotropic limit ($l =m = 0$). For luminal vector GWs, this isotropic part is known to produce a divergence with indefinitely large pulsar distances \cite{Qin:2020hfy, NANOGrav:2021ini, Bernardo:2022rif}, which is tamed in a realistic astrophysical scenario. The divergence however manifests as an unusually large correlation at small angles that renders luminal vector GWs disfavored when compared with observation \cite{NANOGrav:2021ini, Bernardo:2023mxc, Bernardo:2023zna}. For this reason, we go with $v = 0.9$ instead of $v = 1.0$ in the discussion of this section.

\begin{figure}[h!]
\center
	\subfigure[ \ $l = 0$ ]{
		\includegraphics[width = 0.475 \textwidth]{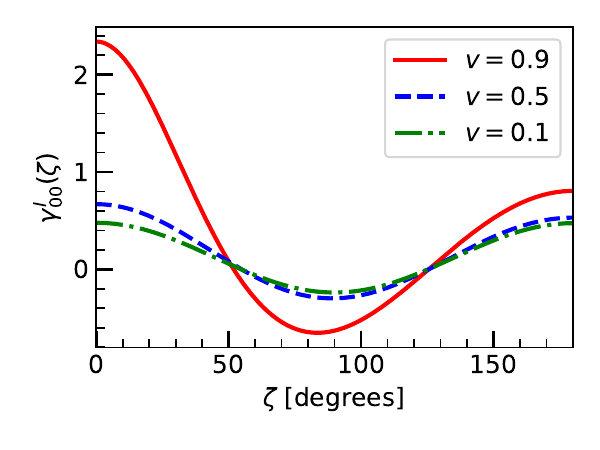}
		}
	\subfigure[ \ $l = 1$ ]{
		\includegraphics[width = 0.475 \textwidth]{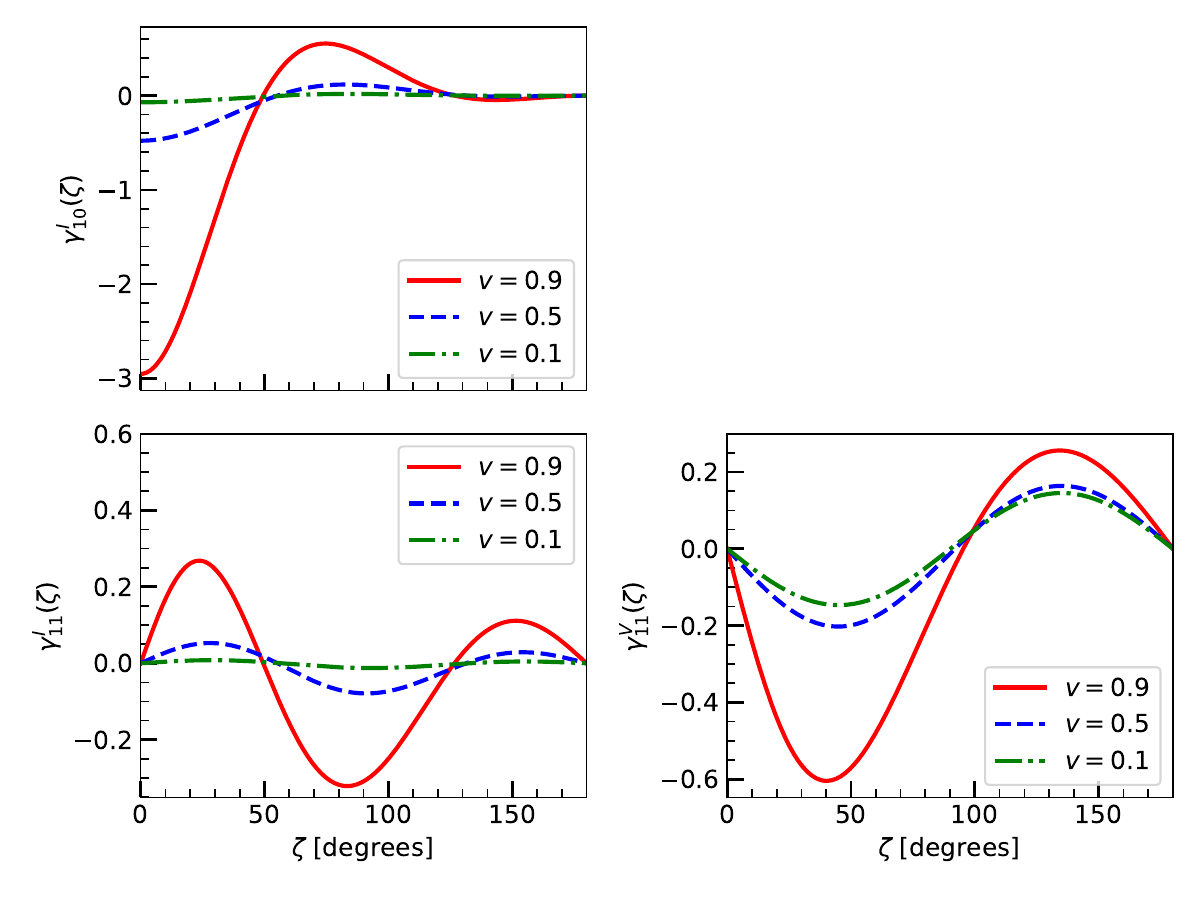}
		}
\caption{(a) Correlations induced by vector modes for $l = m = 0$ (isotropic unpolarized case) and (b) for $l = 1$ for $v = 0.9, 0.5, 0.1$ with $fD = 10^3$.}
\label{fig:gammal01_V}
\end{figure}

\begin{figure}[h!]
    \centering
    \includegraphics[width = 0.95 \textwidth]{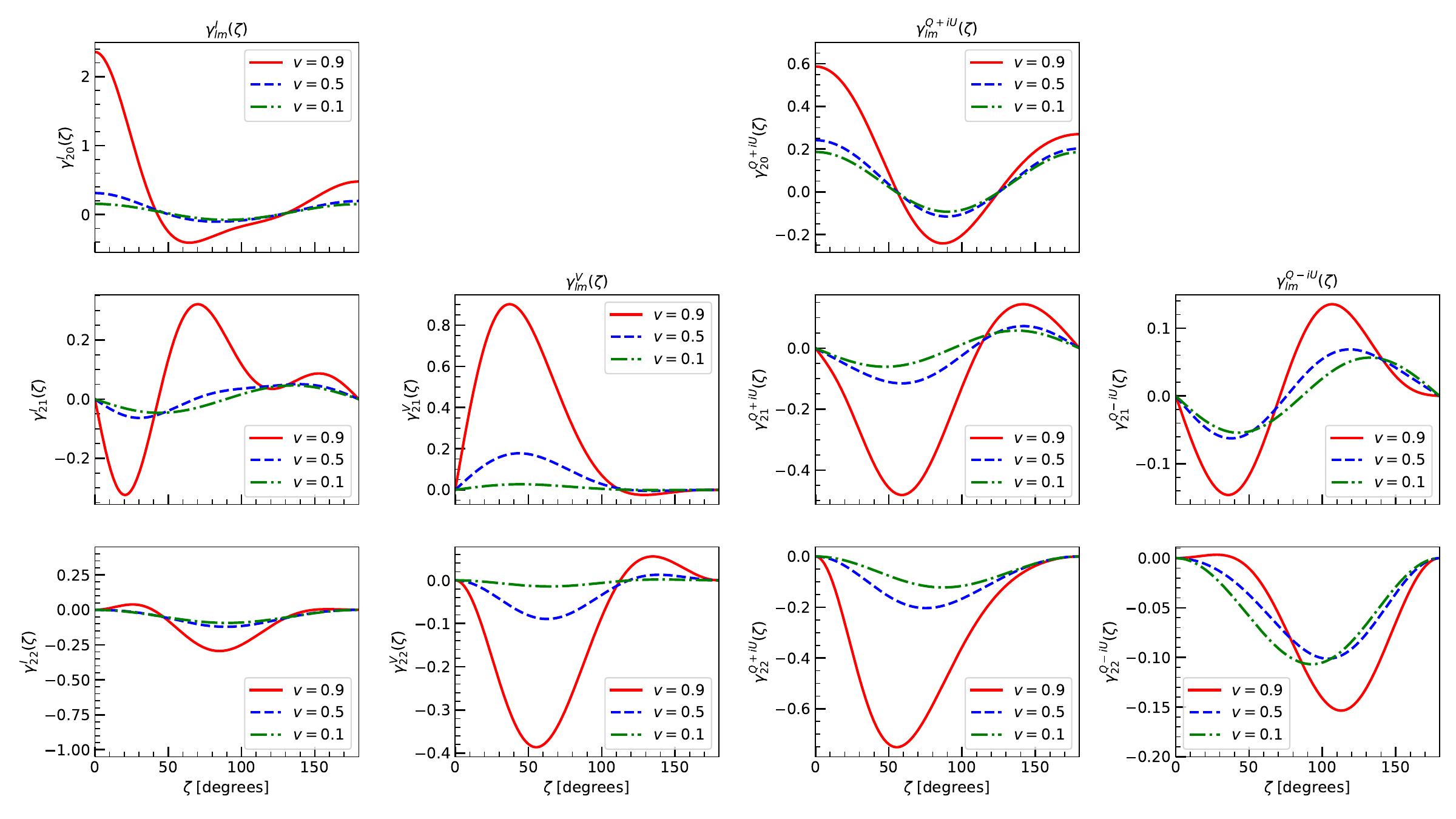}
    \caption{Correlations induced by vector modes for $l = 2$ for $v = 0.9, 0.5, 0.1$ with $fD = 10^3$.}
    \label{fig:gammal2_V}
\end{figure}

The first nontrivial components of anisotropy and polarization of the correlation induced by a stochastic superposition of vector GWs is presented in Figures \ref{fig:gammal01_V} and \ref{fig:gammal2_V}. In this case, we show the anisotropy and polarization modes up to $l \leq 2$ and consider subluminal GW speeds of $v = 0.9, 0.5, 0.1$. This is tied to the spin-1 nature of the vector d.o.f.s that manifest through the Wigner-3j symbol in the correlation. As a consequence, $\gamma_{l0}^V$ vanishes and $\gamma_{lm}^{Q \pm iU}$ returns nontrivial values only beginning at $l = 2$.

It is worth recalling the isotropic limit as a reference \cite{Qin:2020hfy, Bernardo:2022rif}. In this case, the correlation for subluminal vector GWs is dominated by the quadrupole. This explains why from a purely visual standpoint the correlation curves of an isotropic SGWB of vector GWs (Figure \ref{fig:gammal01_V}(a)) may be confused with the tensor ones such as the HD curve (Figure \ref{fig:gammal01_T}(a)). A distinction is that the amount of power is different, which manifests to the size of the correlation and is generally larger for vector GWs because of the extra dipolar contribution. For subluminal vector GWs, the dominant contribution also happens to be the quadrupole. However, compared with tensor modes, the vector GWs next-to-leading order contributions come from a dipole and the octupole of comparable magnitudes at $v \sim 0.5$. This gives the vector-anchored quadrupolar correlation its own flavor, physically distinct from ones produced by tensor modes. At $v \ll 1$ such as $v \sim 0.1$ and the infinite distance limit, the dipolar and the octupolar powers become even more suppressed compared with the quadrupolar contribution. Nonetheless, this turns out to be strongly dependent on the pulsar distance in such a way that it is always possible to distinguish the tensor and vector cases at any GW speeds. We expect that this holds in the astrophysical scenario where the timed millisecond pulsars have $D = {\cal O}(1)\,{\rm kpc}$, though for several with large uncertainties.

Moving on to the anisotropic and polarized components (Figures \ref{fig:gammal01_V}(b) and \ref{fig:gammal2_V}), we realize that subluminal GW propagation in most cases damps it away, reminiscent of the tensor case. This is reflected in Figures \ref{fig:gammal01_V}-\ref{fig:gammal2_V} where the correlation arranged in order of size is more or less in the same order as the GW speed. However, we clarify that this should be seen merely as a rule of thumb as there are clear exceptions particularly with the polarized ones, $\gamma_{2m}^{Q \pm iU}$. Turning this around, these are places in the correlation where it may be worth focusing observations into in order to find signatures of subluminal GWs produced by vector d.o.f.s.

It should be noted that the anisotropy and polarization could be utilized to distinguish between tensor and vector modes in the SGWB. The different powers produced particularly due to the extra dipole of vector modes would be a useful probe of constraining the possible existence of vector GWs. Recall that for tensor GWs, the first nontrivial signals from polarization come at the order $l = 4$. On the other hand, for vector GWS, this leading polarization signature in the correlation appears at $l = 2$. With anisotropy and polarization measurements, we therefore have the extra physical means to understand the nature of gravity through probing the components of the SGWB.

In the end, considering vector GWs may have been impractical to begin with from a production standpoint, as there are not many known astrophysical nor cosmological processes that would be able to give rise to such excitations. The point is however PTAs through correlation measurements of an anisotropic polarized SGWB can be put to good use to independently chip in on this matter of vector GWs.

\subsection{Scalar GW correlations}
\label{subsec:scalargwcorrelations}

Scalar GW polarizations in an isotropic SGWB ($l = m = 0$) manifest as a monopole and a dipole in the correlation, with suppressed quadrupole and higher-order multipoles. This suppression comes about due to the peculiar way that the relativistic scalar modes interfere with each other from the point-of-view of scalar-tensor gravity \cite{Qin:2020hfy, Bernardo:2022vlj, Bernardo:2023pwt}. A physical consequence of this is realized for luminal scalar modes where a potentially divergent longitudinal GW polarization is rendered harmless as it is suppressed by a factor $1 - v^2 \ll 1$ compared to the breathing mode.

\begin{figure}[h!]
    \centering
    \includegraphics[width = 0.975 \textwidth]{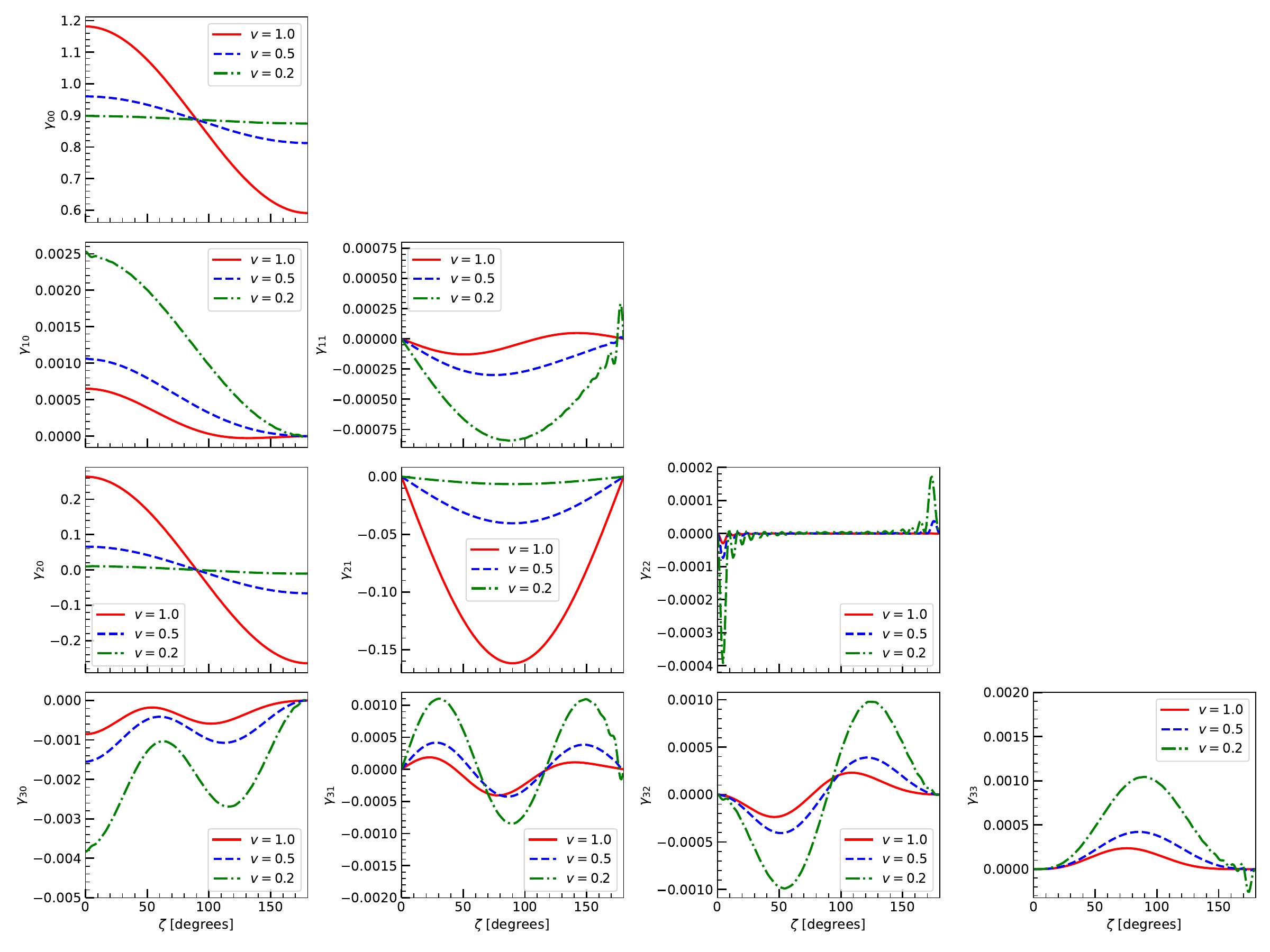}
    \caption{Correlations induced by scalar modes for $l \leq 3$ for $v = 1.0, 0.5, 0.2$ with $fD = 10^3$.}
    \label{fig:gamma_S}
\end{figure}

Figure \ref{fig:gamma_S} shows the various correlations of an SGWB built with gravitational scalar modes. The previous discussion on the isotropic limit is reflected in the top-left corner. In this case, we find that in the luminal limit a dipolar signature in the correlation is visually recognizable. For decreasing values of the speed of the scalar modes, on the other hand, the monopole gradually overtakes the correlation signal, which is shown in the topmost of Figure \ref{fig:gamma_S}. 

We reiterate that for the gravitational scalar d.o.f.s there was no need to break the correlation into Stokes parameters because of the purely longitudinal GW polarization. The rest of Figure \ref{fig:gamma_S} shows the anisotropic components of the correlation of an SGWB  built with scalar GWs. This reveals an interesting result that could be viewed as counter-intuitive: with the exception of the isotropic limit ($l = m = 0$), the correlation curves for scalar GWs manifesting as an SGWB is generally larger for smaller GW speeds. In Figure \ref{fig:gamma_S}, this is exemplified by $\gamma_{1m}$, $\gamma_{22}$, $\gamma_{3m}$. We labelled counter-intuitive based from the previous context of the tensor and vector GWs where the total power drops at subluminal cases. On the other hand, for scalar GWs at subluminal velocities, the monopole becomes dominant but in addition approaching a divergence in the ultra-nonrelativistic limit ($v \ll 1$) \cite{Qin:2020hfy, Bernardo:2022rif}. This is well-known in the isotropic case and now realized at various levels of the anisotropy ($l \geq 1$) through an amplification of the correlation.

It is worth noting that for large pulsar distances the monopole and the dipole behave as $1/v^4$ and $1/v^2$, respectively. Thus, the independent contributions coming from both the breathing and longtudinal GW modes scale as $\sim 1/v^4$ which is sufficient to explain the enhancement of power in the isotropic case at subluminal and ultra-nonrelativistic speeds \cite{Qin:2020hfy, Bernardo:2022rif}. This is suppressed in the isotropic SGWB correlation as shown in Figure \ref{fig:gamma_S}. However, because scalar GWs are nearly-singularly characterized by a monopole and a dipole, the results finally reveal where the enhanced power at small speeds goes into: the anisotropies of the correlation. This could be taken to be constrainable signatures of scalar GW components of the SGWB.

The anisotropic correlation with scalar GWs could be used to independently constrain scalar d.o.f.s in the nanohertz GW regime. A comment is also owed to the tiny high-frequency-wiggles for small GW speeds in several components in Figure \ref{fig:gamma_S} such as $\gamma_{11}$, $\gamma_{22}$, and $\gamma_{33}$. These are in fact sourced by finite pulsar distances and could be ironed out by pulling the pulsars further away to infinity. However, pulsars located at infinity is unrealistic despite the credit to the HD curve which gives a spectacularly accurate analytical solution at large scales. Focusing on the detection of finite pulsar distance effects such as those that manifest at sub-degree angular separations and the ones in Figure \ref{fig:gamma_S} for scalar GWs is a nice addition to PTA science objectives.

\section{Change of basis into bipolar spherical harmonics}
\label{sec:changeofbasis}

It is worthwhile to spend time to tease out the differences of our approach with Ref. \cite{AnilKumar:2023yfw}.

A first is with regard to pulsar terms, which we keep in ours but is dropped in Ref. \cite{AnilKumar:2023yfw}. The dropping of pulsar terms is quite common in the PTA community and for good reason, since PTAs are for the time being mostly sensitive to large angles where pulsar terms are subdominant. Furthermore, this good trade leads to closed-form analytical expressions for the ORF such as the Hellings \& Downs curve. Either way, we show that pulsar terms can be kept, utilizing only the same quantity, $J_l(x)$'s, that are computed for the isotropic case; this implies that extending to the general case of the anisotropic and polarized SGWB requires almost no extra computation. Pulsar terms may be relevant should several millisecond pulsar pairs of arcsecond sky separations be detected, for instance, in the square kilometer array era.

Another key difference, this time on the practical side, is that the Wigner 3j-symbols we mainly use in our expressions are readily available in computational softwares such as scientific computing library `scipy' (in python) that we use in our code \href{https://github.com/reggiebernardo/PTAfast}{PTAfast} \cite{2022ascl.soft11001B}. Bipolar spherical harmonics could of course be computed as this is also heavily used in CMB science in order to study the anisotropy and polarization of the signal. Choosing which basis to express the anisotropy and polarization components is a matter of preference and resources; bipolar spherical harmonics $\{ Y_{l_1}(\hat{e}_a) \otimes Y_{l_2}(\hat{e}_b) \}$ and Wigner 3j-symbols $\left( \begin{array}{ccc}
        l & l_1 & l_2 \\
        m & -m_1 & m_2
    \end{array} \right)$ are related via \cite{Liu:2022skj}
\begin{equation}
\begin{split}
    & \{ Y_{l_1}(\hat{e}_a) \otimes Y_{l_2}(\hat{e}_b) \}_{lm} \\
    & = \sum_{m_1 m_2} (-1)^{m_1} \sqrt{2 l + 1} \left( \begin{array}{ccc}
        l & l_1 & l_2 \\
        m & -m_1 & m_2
    \end{array} \right) Y_{l_1 m_1}(\hat{e}_a) Y^*_{l_2 m_2}(\hat{e}_b) \,.
\end{split}
\end{equation}
Clearly, the spin of the fundamental field that enters through $m$ in our expressions reflect to the same $m$ in the bipolar spherical harmonics in \cite{AnilKumar:2023yfw}, and vice versa.

With that said, a {\it complete} transcription of the expressions in Ref. \cite{AnilKumar:2023yfw} to ours may not possible due to pulsar terms; though, we bear in mind for a future work creating a dictionary that might be able to do this translation with extra terms accounting for pulsar terms. This also has to wait until \cite{AnilKumar:2023yfw} is tested to more cases, i.e., subluminal GWs including non-luminal vector and scalar (breathing and longitudinal) polarizations. Nonetheless, the road has been laid out in \cite{Liu:2022skj} (see Sec. 8). With this, we easily obtain that the $A_{l_1 l_2}^{lm}$'s in equation (10) of \cite{AnilKumar:2023yfw} are related to our $J_l(x)$'s via
\begin{equation}
\begin{split}
    A_{l_1 l_2}^{lm} \propto & \sqrt{(2l_1 + 1) (2l_2 + 1)} J_{l_1}(x) J_{l_2}(y) \\
    & \times \bigg\{ \left[ I_{lm} \left(1 + (-1)^{l+l_1+l_2}\right) +
    V_{lm} \left(1 - (-1)^{l+l_1+l_2}\right) \right] \left( \begin{array}{ccc}
        l & l_1 & l_2 \\
        m & -2 & 2
    \end{array} \right) \\
    & \phantom{gggg} + (Q + iU)_{lm} \left( \begin{array}{ccc}
        l & l_1 & l_2 \\
        -4 & 2 & 2
    \end{array} \right)
    + (Q - iU)_{lm} \left( \begin{array}{ccc}
        l & l_1 & l_2 \\
        4 & -2 & -2
    \end{array} \right) \bigg\} \,
\end{split}
\end{equation}
where $x, y \sim D/\lambda \gg 1$ are ratios between the typical millisecond pulsar distance $D = {\cal O}\left(10^3\right)$ pc and the wavelength of GW $\lambda = {\cal O}\left(10^0\right)$ pc.

We leave out the exact details for future work since this dictionary deserves attention beyond the usual tensor case and luminal GWs.

\section{Conclusions}
\label{sec:conclusions}

To the best of our knowledge, we have derived the most general expressions for the correlation in the PTA and SGWB literature that is valid for subluminal tensor, vector, and scalar GW scenarios and considers finite distances, incorporating often dropped pulsar term modulations relevant at small angular scales. We believe that Section \ref{sec:summary} wraps up our derived formulae in a friendly and codable language that could potentially be used to test the isotropy of the SGWB. Inevitably, this provides a better understanding of the nature of gravity through a grasp of the anisotropic and polarized components of the SGWB with subluminal non-Einsteinian GW polarizations.

However, several issues remain to be settled in a future work. Looking for a middle ground between the formalisms in \cite{Liu:2022skj}---this work's predecessor---and \cite{AnilKumar:2023yfw} is definitely worth considering (Section \ref{sec:changeofbasis}). On top of everything is the notion of the cosmic variance \cite{Allen:2022dzg, Allen:2022ksj, Caliskan:2023cqm, Bernardo:2022xzl}. This is well-understood for the isotropic SGWB in and beyond general relativity, and is invaluable for testing gravity. It remains to generalize this notion and derive it with anisotropy and polarization in the SGWB. On the observational side, there is the question of when the data could be sensitive to the anisotropic and polarized components of correlation. This remains to be addressed. Another future work that is not directly related but is important to the field is to establish a rigorous test of the Gaussianity of the SGWB. This hypothesis is tied to the cosmic variance and the source of the SGWB, and remains to be directly tested with observation.

\acknowledgments

R.C.B. is supported by an appointment to the JRG Program at the APCTP through the Science and Technology Promotion Fund and Lottery Fund of the Korean Government, and was also supported by the Korean Local Governments in Gyeongsangbuk-do Province and Pohang City. This work was supported in part by the National Science and Technology Council (NSTC) of Taiwan, Republic of China, under Grants No. NSTC 112-2123-M-007-001 (G.C.L.) and No. NSTC 112-2112-M-001-067 (K.W.N.).

% \paragraph{Note added.} This is also a good position for notes added
% after the paper has been written.

\appendix
% \section{Some title}
% Please always give a title also for appendices.

\section{A brief history of the overlap reduction function}
\label{sec:orfs_review}

Here are some important results on the ORF of an isotropic SGWB \cite{NANOGrav:2021ini}.

The standard ORF, due to equal parts $+$ and $\times$ GW transverse polarizations, is given by
\begin{equation}
    \Gamma_{ab}^{{\rm T}} \left(\zeta\right) = \Gamma_{ab}^+\left(\zeta\right) + \Gamma_{ab}^\times\left(\zeta\right) \approxeq \dfrac{\delta_{ab}}{2} + C_{ab}\left(\zeta\right) \,,
\end{equation}
where $\zeta$ is the angular separation between a pair of pulsars, $a$ and $b$, and $C_{ab}\left(\zeta\right)$ is the HD curve \cite{Hellings:1983fr},
\begin{equation}
    C_{ab}\left(\zeta\right) = \dfrac{3}{2} \left( \dfrac{1}{3} + \left( \dfrac{1 - \cos \zeta}{2} \right) \left[ \ln \left( \dfrac{1 - \cos \zeta}{2} \right) - \dfrac{1}{6} \right] \right) \,.
\end{equation}
The expression in the parenthesis is the correlation originally derived by Hellings and Downs in 1983. The ORF nowadays is normalized as $\Gamma^{\rm HD}_{ab}(0) = 0.5$. In this context, two physical conditions give rise to the HD correlation: luminal tensor modes and pulsars at infinity.

There are also phenomenological ORFs that are not necessarily due to gravitational degrees of freedom, but nonetheless are worth considering given their comparatively large signal-to-noise ratio in the previous iterations of the data \cite{NANOGrav:2021ini}. These are the GW-like monopole and dipole:
\begin{equation}
    \Gamma_{ab}^{\text{GW-mon}} = \dfrac{\delta_{ab}}{2} + \dfrac{1}{2} \,,
\end{equation}
and
\begin{equation}
    \Gamma_{ab}^{\text{GW-dip}} = \dfrac{\delta_{ab}}{2} + \dfrac{\cos \zeta}{2} \,.
\end{equation}
The `GW-like' monicker is owed to the first term, $\delta_{ab}/2$, put in by hand, to compensate possible unaccounted for small scale power ($a = b$) irrelevant for the large scale correlation ($a \neq b$). Without the first term, these are simply referred to as monopole and dipole correlations.

Several ORFs have also been derived for non-Einsteinian GW polarizations when the modes are on the light cone and the pulsars are at infinity. The `breathing' mode or scalar transverse polarization gives rise to the correlation
\cite{Chamberlin:2011ev}
\begin{equation}
    \Gamma_{ab}^{{\rm b}} \approx \dfrac{\delta_{ab}}{2} + \dfrac{ 3 + \cos \zeta }{8} \,.
\end{equation}
On the other hand, a related `longitudinal' scalar mode correlation, $\Gamma_{aa}^l$, diverges in the same limit. Keeping the scalar modes on the lightcone with a GW frequency $f$ but with pulsars at a finite distance $D$, then in the limit $fD \gg 1$, it can be shown that the autocorrelation function of the scalar longitudinal mode behaves as \cite{Chamberlin:2011ev}
\begin{equation}
     \Gamma_{aa}^{\rm l} \sim \dfrac{3\pi^2}{4} fD - 3 \ln \left( 4\pi fD \right) + \dfrac{37}{8} - 3 \gamma_\text{E} \,,
\end{equation}
where $\gamma_\text{E}$ is Euler's constant. A similar situation arises with vector modes; in which case, their ORF can be found to be
\begin{equation}
    \Gamma_{ab}^\text{V} = \Gamma_{ab}^{\rm x} + \Gamma_{ab}^{\rm y} \approx 3 \log \left( \dfrac{2}{1 - \cos \zeta} \right) - 4 \cos \zeta - 3 \,,
\end{equation}
except that the autocorrelation diverges,
\begin{equation}
    \Gamma_{aa}^\text{V} \sim 6 \ln \left( 4 \pi fD \right) - 14 + 6 \gamma_\text{E} \,,
\end{equation}
in the infinite distance limit.

It must be said that the divergences do not manifest in practice as the pulsars are always at finite distances in a finite volume of the observable Universe. Analysis of the results for finite pulsar distances attest to this.

\section{Triple spherical harmonics integral and the Wigner-3j symbol}
\label{sec:3Y3j}

We lay down identities on spin weighted spherical harmonics $\,_s Y_{lm}\left(\hat{n}\right)$ relevant in the body of the paper.

For $s = 0$, $\,_s Y_{lm}\left(\hat{n}\right)$ reduces to the spherical harmonic $ Y_{lm}\left(\hat{n}\right)$. In general, $\,_s Y_{lm}\left(\hat{n}\right)$ satisfies the orthogonality relation
\begin{equation}
    \int_{S^2} d\hat{n} \,_s Y_{lm}^*\left(\hat{n}\right) \,_s Y_{l'm'}\left(\hat{n}\right) = \delta_{ll'} \delta_{mm'}
\end{equation}
and completeness relation
\begin{equation}
    \sum_{lm} \,_s Y^*_{lm}\left(\hat{n}\right) \,_s Y_{lm}\left(\hat{n}' \right) = \delta^{(2)}\left( \hat{n} - \hat{n}' \right) = \delta\left(\phi - \phi'\right) \delta\left(\cos \theta - \cos \theta' \right) \,.
\end{equation}
Spin-weighted spherical harmonics satisfy the conjugate identity
\begin{equation}
    \,_s Y_{lm}^*\left( \hat{n} \right) = (-1)^{s + m} \,_{-s}Y_{l-m}\left( \hat{n} \right) \,,
\end{equation}
and a triple spherical harmonics identity
\begin{equation}
\begin{split}
    \int d \hat{e} \,_{s_1} Y_{l_1 m_1}\left(\hat{e}\right) \,_{s_2}Y_{l_2 m_2}\left(\hat{e}\right) \,_{s_3}Y_{l_3 m_3}\left(\hat{e}\right) =
    & \sqrt{\dfrac{(2l_1+1)(2l_2+1)(2l_3+1)}{4\pi}} \\
    & \ \ \times \left( \begin{array}{ccc}
    l_1 & l_2 & l_3 \\
    -s_1 & -s_2 & -s_3
    \end{array} \right)
    \left( \begin{array}{ccc}
    l_1 & l_2 & l_3 \\
    m_1 & m_2 & m_3
    \end{array} \right)
\end{split}
\end{equation}
where $\left(\begin{array}{ccc} a & b & c \\ d & e & f \end{array}\right)$ is the Wigner-3j symbol. The 3j symbol vanishes unless $|l_1 - l_2| < l_3 < l_1 + l_2$ and $m_1 + m_2 + m_3 = 0$; if $m_1 = m_2 = m_3 = 0$, then $l_1 + l_2 + l_3$ must be an even integer; the 3j symbol satisfies a reflection property
\begin{equation}
    \left( \begin{array}{ccc}
    l_1 & l_2 & l_3 \\
    m_1 & m_2 & m_3
    \end{array} \right)
    = (-1)^{l_1 + l_2 + l_3}
    \left( \begin{array}{ccc}
    l_1 & l_2 & l_3 \\
    -m_1 & -m_2 & -m_3
    \end{array} \right) \,.
\end{equation}

\section{GW polarizations}
\label{sec:gwpolarizations}

For a GW propagating along the $\hat{\Omega}$ direction, the polarization basis for tensor, vector, and scalar GWs can be expressed as \cite{Boitier:2021rmb}
\begin{equation}
    \varepsilon^{+} = \hat{m} \otimes \hat{m} - \hat{n} \otimes \hat{n} \phantom{gggg} {\rm and} \phantom{gggg} \varepsilon^{\times} = \hat{m} \otimes \hat{n} + \hat{n} \otimes \hat{m} \,,
\end{equation}
\begin{equation}
    \varepsilon^{x} = \hat{m} \otimes \hat{\Omega} + \hat{\Omega} \otimes \hat{m} \phantom{gggg} {\rm and} \phantom{gggg} \varepsilon^{y} = \hat{n} \otimes \hat{\Omega} + \hat{\Omega} \otimes \hat{n} \,,
\end{equation}
\begin{equation}
    \varepsilon^{b} = \hat{m} \otimes \hat{m} + \hat{n} \otimes \hat{n} \phantom{gggg} {\rm and} \phantom{gggg} \varepsilon^{l} = \sqrt{2} \hat{\Omega} \otimes \hat{\Omega} \,,
\end{equation}
respectively, where $\left( \hat{m}, \hat{n}, \hat{\Omega} \right)$ form an orthonormal basis. $\left(\varepsilon^{+}, \varepsilon^{\times} \right)$ are the familiar transverse-traceless tensor GW polarizations, $\left(\varepsilon^{x}, \varepsilon^{y} \right)$ are the corresponding vector GW polarizations, and $\left(\varepsilon^{b}, \varepsilon^{l} \right)$ for scalar GWs.

It is useful to point the GW/$\hat{\Omega}$ along the $z$-direction such that the orthonormal basis $\left( \hat{m}, \hat{n}, \hat{\Omega} \right)$ may be written as
\begin{eqnarray}
\hat{m} &=& \cos \varphi \ \hat{x} + \sin \varphi \ \hat{y} \\
\hat{n} &=& -\sin \varphi \ \hat{x} + \cos \varphi \ \hat{y} \\
\hat{\Omega} &=& \hat{z} \,.
\end{eqnarray}
Calculations can be generalized at the end by a three-dimensional rotation of $\hat{\Omega}$. With the Cartesian basis $\left( \hat{x}, \hat{y}, \hat{z} \right)$, the GW polarization basis can be identified to be
\begin{equation}
    \varepsilon^{+} = 
    \left(
    \begin{array}{ccc}
    \cos(2\varphi) & \sin(2\varphi) & 0 \\
    \sin(2\varphi) & -\cos(2\varphi) & 0 \\
    0 & 0 & 0 
    \end{array}
    \right) \phantom{gggg} {\rm and} \phantom{gggg} \varepsilon^{\times} = 
    \left(
    \begin{array}{ccc}
    -\sin(2\varphi) & \cos(2\varphi) & 0 \\
    \cos(2\varphi) & -\sin(2\varphi) & 0 \\
    0 & 0 & 0 
    \end{array}
    \right) \,,
\end{equation}
\begin{equation}
    \varepsilon^{x} = 
    \left(
    \begin{array}{ccc}
    0 & 0 & \cos \varphi \\
    0 & 0 & \sin \varphi \\
    \cos \varphi & \sin \varphi & 0 
    \end{array}
    \right) \phantom{gggg} {\rm and} \phantom{gggg} \varepsilon^{y} = 
    \left(
    \begin{array}{ccc}
    0 & 0 & -\sin \varphi \\
    0 & 0 & \cos \varphi \\
    -\sin \varphi & \cos \varphi & 0 
    \end{array}
    \right) \,,
\end{equation}
\begin{equation}
    \varepsilon^{b} = 
    \left(
    \begin{array}{ccc}
    1 & 0 & 0 \\
    0 & 1 & 0 \\
    0 & 0 & 0 
    \end{array}
    \right) \phantom{gggg} {\rm and} \phantom{gggg} \varepsilon^{l} = 
    \sqrt{2} \left(
    \begin{array}{ccc}
    0 & 0 & 0 \\
    0 & 0 & 0 \\
    0 & 0 & 1 
    \end{array}
    \right) \,.
\end{equation}

\section{Calculation of the $J_{lm}(fD, \hat{k})$'s}
\label{sec:Jlmcalculations}

This appendix lays down the detailed calculation of the $J_{lm}(fD, \hat{k})$'s.

\subsection{Tensor GW modes}
\label{subsec:Jlmtensor}

To illustrate our strategy, we consider tensor GWs: we fix $\hat{k} = \hat{z}$; then, rotate $\hat{z} \rightarrow \hat{k}$ to obtain $J_{lm}\left(fD, \hat{k} \right)$.

We define the right- and left-handed complex circular polarization basis tensors:
\begin{equation}
    \varepsilon^R = \dfrac{\varepsilon^+ + i \varepsilon^\times}{\sqrt{2}} \ \ \ \ \ \text{and} \ \ \ \ \varepsilon^L = \dfrac{\varepsilon^+ - i \varepsilon^\times}{\sqrt{2}} \,.
\end{equation}
The contraction of the detector tensor, $d^{ij}$, with the basis tensors give
\begin{equation}
    d^{ij} \varepsilon_{ij}^{R, L} = \sqrt{\dfrac{16\pi}{15}} Y_{2 \pm 2} \left( \hat{e} \right) \,,
\end{equation}
where the helicity $R$ ($L$) takes $m = + 2$ ($-2$). Substituting this into \eqref{eq:Jlm_def} with $Y_{LM}\left(\hat{k} = \hat{z} \right) = \delta_{M0} \sqrt{(2L+1)/4\pi}$, we obtain
\begin{equation}
    J_{lm}^{R,L} \left( fD, \hat{k} \right) = \int_0^{2\pi f D v} \dfrac{d x}{v} \ e^{i x/v} \sum_{L} 4 \pi i^L \sqrt{\dfrac{2L+1}{15}} j_L(x) \int_{S^2} d\hat{e} \ Y_{2\pm2}\left(\hat{e}\right) Y_{L0}\left( \hat{e} \right) Y_{lm}^*\left(\hat{e}\right) \,.
\end{equation}
The triple spherical harmonics integral vanishes; except with the following nontrivial values for $m = \pm 2$ and $L = l - 2, l, l + 2$:
\begin{equation}
    \int_{S^2} d\hat{e} \ Y_{2\pm 2}\left(\hat{e}\right) Y_{(l - 2) 0}\left(\hat{e}\right) Y^*_{l \pm 2}\left(\hat{e}\right) = \sqrt{\dfrac{15}{32\pi}} \left( \dfrac{(l-1)l(l+1)(l+2)}{(2l-3)(2l-1)^2(2l+1)} \right)^{1/2} \,,
\end{equation}
\begin{equation}
    \int_{S^2} d\hat{e} \ Y_{2\pm 2}\left(\hat{e}\right) Y_{l0}\left(\hat{e}\right) Y^*_{l \pm 2}\left(\hat{e}\right) = - \sqrt{\dfrac{15}{8\pi}} \left( \dfrac{(l-1)l(l+1)(l+2)}{(2l-1)^2(2l+3)^3} \right)^{1/2} \,,
\end{equation}
and
\begin{equation}
    \int_{S^2} d\hat{e} \ Y_{2\pm 2}\left(\hat{e}\right) Y_{(l + 2) 0}\left(\hat{e}\right) Y^*_{l \pm 2}\left(\hat{e}\right) = \sqrt{\dfrac{15}{32\pi}} \left( \dfrac{
(l-1)l(l+1)(l+2)}{(2l+1)(2l+3)^2(2l+5)} \right)^{1/2} \,.
\end{equation}
The last expression reduces to
\begin{equation}
\begin{split}
    J_{lm}^{R,L}\left(fD, \hat{z}\right) = 
    & -\delta_{m\pm 2} 2 \pi i^l \sqrt{ \dfrac{2l+1}{8\pi} \dfrac{(l + 2)!}{(l - 2)!} } \int_0^{2\pi fDv} \dfrac{dx}{v} \ e^{ix/v} \\
    & \ \ \times \left( \dfrac{j_{l-2}(x)}{(2l-1)(2l+1)} + \dfrac{2j_l(x)}{(2l-1)(2l+3)} + \dfrac{j_{l+2}(x)}{(2l+1)(2l+3)} \right) \,.
\end{split}
\end{equation}
Utilizing the spherical Bessel function recursion relation
\begin{equation}
\label{eq:bessel_id1}
    \dfrac{j_l(x)}{x} = \dfrac{j_{l-1}(x) + j_{l+1}(x)}{2l+1} \,,
\end{equation}
we obtain
\begin{equation}
    J_{lm}^{R,L}\left(fD, \hat{z}\right) = - \delta_{m\pm 2} 2\pi i^l \sqrt{ \dfrac{2l+1}{8\pi} \dfrac{(l + 2)!}{(l - 2)!} } \int_0^{2\pi fDv} \dfrac{dx}{v} \ e^{ix/v} \dfrac{j_l(x)}{x^2} \,.
\end{equation}

Then, the result is obtained by rotating $\hat{z}$ to $\hat{k} = (\theta,\phi)$ direction via the general rule
\begin{equation}
    J^A_{lm}\left( fD, \hat{k} \right) = \sum_{m'} D_{m'm}^{l*} \left(-\alpha,-\theta,-\phi\right) J^A_{lm'}\left( fD, \hat{z} \right) \,,
\end{equation}
where $D_{m'm}^l(-\alpha,-\theta,-\phi)$ is the Wigner-D matrix given by
\begin{equation}
    D_{m'm}^l(-\alpha,-\theta,-\phi) = \sqrt{\dfrac{4\pi}{2l + 1}} \, _{-m'}Y_{lm}\left(\theta,\phi\right) e^{i m' \alpha} \,,
\end{equation}
and $\,_s Y_{lm}\left(\hat{e}\right)$'s are spin weighted spherical harmonics (Appendix \ref{sec:3Y3j}). Our final result is given by \eqref{eq:Jlm_tensor}.

\subsection{Vector GW modes}
\label{subsec:Jlmvector}

As with tensor GWs, we rely on right- and left-handed helicity basis tensors to deal with vector GWs:
\begin{equation}
    \varepsilon^{VR} = \dfrac{\varepsilon^x + i \varepsilon^y}{\sqrt{2}} \ \ \ \ \ \text{and} \ \ \ \ \varepsilon^{VL} = \dfrac{\varepsilon^x - i \varepsilon^y}{\sqrt{2}} \,.
\end{equation}
Similarly, the contraction of the detector tensor, $d^{ij}$, with the basis tensors gives
\begin{equation}
    d^{ij} \varepsilon_{ij}^{VR, VL} = \mp \sqrt{\dfrac{16\pi}{15}} Y_{2 \pm 1} \left( \hat{e} \right) \,,
\end{equation}
where the upper (lower) signs correspond to $VR$ ($VL$). In this case, the triple spherical harmonics integration reduces to
\begin{equation}
\small
    \int d\hat{e} \ Y_{21}\left(\hat{e}\right) Y_{L0}\left(\hat{e}\right) Y_{lm}\left(\hat{e}\right) = \sqrt{\dfrac{15}{2 \pi }} \dfrac{ (-l+L+1) \sqrt{l (l+1) (2 l+1) (2 L+1)} \left(-(l-1) (l+2)+L^2+L\right)}{(-l+L-2) (l+L-1) (l+L+1) (l+L+3) (l-L)! (-l+L+2)!}
\end{equation}
which holds for $m = -1, l \geq 1, l - 2 \leq L \leq l + 2$ and $L + l \geq 2$, and
\begin{equation}
    \int d\hat{e} \ Y_{2-1}\left(\hat{e}\right) Y_{L0}\left(\hat{e}\right) Y_{lm}\left(\hat{e}\right) = (-1)^{L + l} \int d\hat{e} \ Y_{21}\left(\hat{e}\right) Y_{L0}\left(\hat{e}\right) Y_{lm}\left(\hat{e}\right)
\end{equation}
which holds for $m = 1, l \geq 1, l - 2 \leq L \leq l + 2$ and $L + l \geq 2$. Since $L + l$ is even, the two integrals become equal except with $m = \mp 1$. The integral is then more compactly expressed as
\begin{equation}
\begin{split}
    \int d\hat{e} \ Y_{2\pm 1}\left(\hat{e}\right) Y_{L0}\left(\hat{e}\right) Y_{lm}\left(\hat{e}\right) = \ & \delta_{m \mp 1}
    \bigg[ \delta_{l1} \delta_{L1} \left( - \sqrt{\dfrac{3}{20\pi}} \right) + \delta_{l1} \delta_{L3} \sqrt{ \dfrac{9}{140\pi} } \\
    & + \Theta\left( l - 2 \right) \bigg[
    \delta_{L(l-2)} \left( - \sqrt{\dfrac{15}{2 \pi }} \dfrac{(l-1) \sqrt{l \left(4 l^3-7 l-3\right)}}{2 (2 l-3) (2 l-1) (2 l+1)} \right)
    \\
    & \phantom{ggggggggggg} + \delta_{Ll} \left( - \sqrt{\dfrac{15}{2 \pi }} \dfrac{ \sqrt{l (l+1) (2 l+1)^2}}{2 (2 l-1) (2 l+1) (2 l+3)} \right) \\
    & \phantom{} + \delta_{L(l+2)} \left( \sqrt{\dfrac{15}{2 \pi }} \dfrac{ l (l+1) (l+2)}{2 (2 l+3) \sqrt{l (l+1) (2 l+1) (2 l+5)}} \right) \bigg] \bigg] \,,
\end{split}
\end{equation}
where $\Theta(x)$ is the step function. Substituting this into \eqref{eq:Jlm_def}, we obtain the simplified expression
\begin{equation}
\begin{split}
    J_{lm}^{VR,VL} \left( fD, \hat{z} \right) = & \mp \sqrt{ \dfrac{16\pi}{15} } \delta_{m \pm 1} \bigg[ - \delta_{l1} \dfrac{3i}{2 \sqrt{5}} \int_0^{2\pi f D v} \dfrac{dx}{v} \ e^{ix/v} \left(j_1(x) + j_3(x)\right) \\
    & \ \ + \Theta(l - 2) \sqrt{\dfrac{15}{2}} \dfrac{i^l}{2} \sqrt{l (l + 1)(2l + 1)} \int_0^{2\pi f D v} \dfrac{dx}{v} \ e^{ix/v} \dfrac{d}{dx} \left( \dfrac{j_l(x)}{x} \right) \bigg] \,.
\end{split}
\end{equation}
Notably, the $l = 1$ piece can be continued to the same form as the $l \geq 2$ contributions. Thus, the expression becomes
\begin{equation}
\begin{split}
    J_{(l \geq 1)m}^{VR,VL} \left( fD, \hat{z} \right) = \mp \delta_{m \pm 1} \sqrt{2\pi} i^l \sqrt{l (l + 1)(2l + 1)} \int_0^{2\pi f D v} \dfrac{dx}{v} \ e^{ix/v} \dfrac{d}{dx} \left( \dfrac{j_l(x)}{x} \right) \,.
\end{split}
\end{equation}
Rotating $\hat{z}$ to an arbitrary $\hat{k}$ direction, we obtain the final result \eqref{eq:Jlm_vector}.

\subsection{Scalar GW modes}
\label{subsec:Jlmscalar}

We follow the same strategy to calculate $J_{lm}^{A} \left(fD, \hat{k}\right)$ for scalar GWs. The difference is that we calculate the correlations directly for the scalar transverse and longitudinal modes; a right- and left- handed helicity basis is not suitable since one of the independent scalar GW polarizations is purely longitudinal.

This time, the contraction of the detector tensor, $d^{ij}$, with the polarization basis for the ST and SL modes become
\begin{equation}
\label{eq:deps_ST}
    d^{ij} \varepsilon_{ij}^{\text{ST}} = \sin^2 \theta = \dfrac{4\sqrt{\pi}}{3} Y_{00}\left(\hat{e}\right) - \dfrac{4}{3} \sqrt{\dfrac{\pi}{5}} Y_{20}\left(\hat{e}\right)
\end{equation}
and
\begin{equation}
\label{eq:deps_SL}
    d^{ij} \varepsilon_{ij}^{\text{SL}} = \sqrt{2} \cos^2 \theta = \sqrt{2} \left( \dfrac{2\sqrt{\pi}}{3} Y_{00}\left(\hat{e}\right) + \dfrac{4}{3} \sqrt{\dfrac{\pi}{5}} Y_{20}\left(\hat{e}\right) \right) \,.
\end{equation}
We proceed with the summation and integration in \eqref{eq:Jlm_def} independently for the scalar GW polarizations.

In the ST case, the only surviving term in the sum over ${L,M}$ in \eqref{eq:Jlm_def} is $M = 0$. The relevant triple spherical harmonics integrals are
\begin{equation}
\label{eq:IntY3_Scalar1}
    \int d \hat{e} \ Y_{00}\left(\hat{e}\right) Y_{L0}\left(\hat{e}\right) Y_{l m}\left(\hat{e}\right) = \dfrac{\delta_{m0}\delta_{lL}}{\sqrt{4\pi}} 
\end{equation}
and
\begin{equation}
\label{eq:IntY3_Scalar2}
\begin{split}
    & \int d \hat{e} \ Y_{20}\left(\hat{e}\right) Y_{L0}\left(\hat{e}\right) Y_{l m}\left(\hat{e}\right) \\
    & \ \ = \dfrac{ \delta_{m0} }{2} \sqrt{\dfrac{5}{\pi}} \dfrac{(L-l +1)^2 (L+l ) (L+l +2) \sqrt{(2 L+1) (2 l +1)} \Gamma (-L+l +3)}{(-L+l +2)^2 (L+l -1) (L+l +1) (L+l +3) \Gamma\left(l -L+1\right)^2 \Gamma (L-l +3)}
\end{split}
\end{equation}
where \eqref{eq:IntY3_Scalar2} holds if $l - 2 \leq L \leq l + 2$ and $L + l \geq 2$; otherwise, it vanishes. In summing $L$, we note that $l_1 + l_2 + l_3$ in the $3j$ symbol must be an even integer since $m_1 = m_2 = 0$. By the end, this leaves three terms corresponding to $L = l - 2, l, l + 2$. We also pull out $l = 0, 1$ terms from the $L = l + 2$ contribution, and the $l = 1$ from the $L = l$ contribution so that we are able to add the terms for $l \geq 2$ coming from all $L = l - 2, l, l + 2$ terms. In this way, the last integral becomes
\begin{equation}
\label{eq:IntY3_Scalar2b}
    \int d \hat{e} \ Y_{20}\left(\hat{e}\right) Y_{L0}\left(\hat{e}\right) Y_{l m}\left(\hat{e}\right) =
    \begin{cases}
    \dfrac{3\delta_{m0}}{4} \sqrt{\dfrac{5}{\pi }} \dfrac{ (l-1) l}{ \sqrt{2 l-3} (2 l-1) \sqrt{2 l+1}} &, \ \ \ \ L = l-2 , l \geq 2 \phantom{\dfrac{\dfrac{1}{2}}{\dfrac{1}{2}}} \\
    \delta_{m0} \sqrt{\dfrac{5}{\pi }} \dfrac{ l (l+1)}{2(2l - 1)(2l+3)} &, \ \ \ \ L = l , l \geq 1 \phantom{\dfrac{\dfrac{1}{2}}{\dfrac{1}{2}}} \\
    \dfrac{3\delta_{m0}}{4}\sqrt{\dfrac{5}{\pi }} \dfrac{ (l+1) (l+2)}{(2 l+3) \sqrt{(2 l+1) (2 l+5)}} &, \ \ \ \ L = l+2 , l \geq 0 \phantom{\dfrac{\dfrac{1}{2}}{\dfrac{1}{2}}} \,,
    \end{cases} 
\end{equation}
and we obtain the simplified expression
\begin{equation}
\label{eq:J00_STc}
\begin{split}
    J_{lm}^\text{ST} \left( fD, \hat{z} \right)
    = \ & \delta_{m0} \delta_{l0} \dfrac{2\sqrt{ \pi}}{3} \int_0^{2\pi fDv} \dfrac{d x}{v} \ e^{i x/v} \left( j_0(x) + j_2(x) \right) \\
    & + \delta_{m0}\delta_{l1} \dfrac{2 \sqrt{3\pi} i}{5}\int_0^{2\pi fDv} \dfrac{d x}{v} \ e^{i x/v} \left( j_1(x) + j_3(x) \right) \\
    - \delta_{m0} & \Theta\left(l - 2\right) 4\pi i^l \sqrt{\dfrac{2l+1}{4\pi}} \int_0^{2\pi fDv} \dfrac{d x}{v} \ e^{i x/v} \left[ \dfrac{d}{dx} \left( \dfrac{j_l(x)}{x} \right) -\dfrac{(l-1)(l+2)}{2} \dfrac{j_l(x)}{x^2} \right] \,.
\end{split}
\end{equation}
Making good use of the spherical Bessel function differential equation,
\begin{equation}
x^2 j_l''(x) + 2xj_l'(x) + \left( x^2 - l(l+1) \right) j_l(x) = 0 \,,    
\end{equation}
and \eqref{eq:bessel_id1}, we obtain the further simplification
\begin{equation}
\label{eq:J00_STd}
\begin{split}
    J_{lm}^\text{ST} \left( fD, \hat{z} \right)
    = \delta_{m0} 2\pi i^l \sqrt{\dfrac{2l+1}{4\pi}} \int_0^{2\pi fDv} \dfrac{d x}{v} \ e^{i x/v} \left( j_l''(x) + j_l(x) \right) \,.
\end{split}
\end{equation}
Performing the three-dimensional rotation, as we did with the tensor and vector cases, we end up with the final expression \eqref{eq:JlmSTfinal}.

Similarly, for the SL modes, summing over $L$ and using the spherical harmonics integral identities, we obtain
\begin{equation}
\label{eq:J00_SLc}
\begin{split}
    \dfrac{J_{lm}^\text{SL} \left( fD, \hat{z} \right)}{\sqrt{2}}
    = \ & \delta_{m0} \delta_{l0} \dfrac{2\sqrt{ \pi}}{3} \int_0^{2\pi fDv} \dfrac{d x}{v} \ e^{i x/v} \left( \dfrac{j_0(x)}{2} - j_2(x) \right) \\
    & + \delta_{m0}\delta_{l1} \dfrac{2 \sqrt{3\pi} i}{5}\int_0^{2\pi fDv} \dfrac{d x}{v} \ e^{i x/v} \left( \dfrac{3 j_1(x)}{2} - j_3(x) \right) \\
    + \delta_{m0} \Theta\left(l - 2\right) & 4\pi i^l \sqrt{\dfrac{2l+1}{4\pi}} \int_0^{2\pi fDv} \dfrac{d x}{v} \ e^{i x/v} \left[ \dfrac{d}{dx} \left( \dfrac{j_l(x)}{x} \right) - \dfrac{(l-1)(l+2)}{2}  \dfrac{j_l(x)}{x^2} +  \dfrac{j_l(x)}{2} \right] \,.
\end{split}
\end{equation}
Then, utilizing the spherical Bessel function differential equation and recursion relation, we get the simplification
\begin{equation}
\label{eq:J00_SLd}
\begin{split}
    \dfrac{J_{lm}^\text{SL} \left( fD, \hat{z} \right)}{\sqrt{2}}
    = - \delta_{m0} 2\pi i^l \sqrt{\dfrac{2l+1}{4\pi}} \int_0^{2\pi fDv} \dfrac{d x}{v} \ e^{i x/v} j_l''\left(x\right) \,.
\end{split}
\end{equation}
Performing the three-dimensional rotation leads to the final result \eqref{eq:JlmSLfinal}.

%\bibliographystyle{JHEP}
%\bibliography{refs}

\providecommand{\href}[2]{#2}\begingroup\raggedright\endgroup

\end{document}